\documentclass[11pt]{article}
\pdfoutput=1
\usepackage{fullpage}
\usepackage{fancyhdr} % to change header/footer of any page
\usepackage{cite}     % to cite references
\usepackage[numbers]{natbib}
\usepackage[top=2.5cm, bottom=2.5cm, left=2.5cm, right=2.5cm]{geometry}
\usepackage{systeme}
\usepackage{epsfig}   % to include eps figures

\usepackage{graphicx} % to manage external pictures
\usepackage{epstopdf}
\usepackage{amsmath}  % advanced math extensions
\usepackage{amssymb}  % adds new symbols in math mode
\usepackage{amsfonts} % set of misc. tex fonts
\usepackage{geometry}
\usepackage{authblk}
\usepackage{mathtools}
\usepackage{amsthm}
\usepackage{siunitx}
\usepackage{booktabs}
\usepackage{multirow}
\usepackage{amsthm}
{

      \theoremstyle{break}
      \newtheorem{assumption}{Assumption}
      \theoremstyle{break}
      \newtheorem{lemma}{Lemma}
      \theoremstyle{break}
      \newtheorem{definition}{Definition}
      \theoremstyle{break}
      
      \theoremstyle{plain}
      
      \theoremstyle{plain}
      
      \theoremstyle{break}
      
      \theoremstyle{break}
      \newtheorem{claim}{Claim}
      \theoremstyle{break}
      
      \theoremstyle{break}
      \newtheorem{consistency}{Consistency}
}
\usepackage{verbatim}
\usepackage{hyperref}

\usepackage{cleveref}
\usepackage{morefloats}
\usepackage{slashbox}
\usepackage{color}
\usepackage{pbox}
\usepackage{titlesec}

\usepackage{booktabs,dcolumn,caption}
\newcolumntype{d}[1]{D{.}{.}{#1}} % "decimal" column type
\newcolumntype{d}[1]{D..{#1}}

\titleformat*{\section}{\Large\bfseries}
\titleformat*{\subsection}{\large\bfseries}
\titleformat*{\subsubsection}{\large\bfseries}
\newcommand{\defeq}{\vcentcolon=}

\newcommand{\var}{\mathbf{var\,}}

\date{\today}
\title{\textsc{Random matrix approach to estimation of high-dimensional factor models}}
\author[1]{Joongyeub Yeo\thanks{uriyeobi@gmail.com, Institute for Computational and Mathematical Engineering, Stanford University, Stanford, CA 94305, USA}}
\author[2]{George Papanicolaou\thanks{papanico@math.stanford.edu, Department of Mathematics, Stanford University, Stanford, CA 94305, USA}}
\affil[1]{Institute for Computational and Mathematical Engineering, Stanford University}
\affil[2]{Department of Mathematics, Stanford University}
\begin{document}
% make titlec
\maketitle
%\tableofcontents
%\listoffigures
%\vspace{10 mm}
% sections with numbering.

\section*{Abstract}
In dealing with high-dimensional data sets, factor models are often useful for dimension reduction. The estimation of factor models has been actively studied in various fields. In the first part of this paper, we present a new approach to estimate high-dimensional factor models, using the empirical spectral density of residuals. The spectrum of covariance matrices from financial data typically exhibits two characteristic aspects: a few spikes and bulk. The former represent factors that mainly drive the features and the latter arises from idiosyncratic noise. Motivated by these two aspects, we consider a minimum distance between two spectrums; one from a covariance structure model and the other from real residuals of financial data that are obtained by subtracting principal components. Our method simultaneously provides estimators of the number of factors and information about correlation structures in residuals. Using free random variable techniques, the proposed algorithm can be implemented and controlled  effectively. Monte Carlo simulations confirm that our method is robust to noise or the presence of weak factors. Furthermore, the application to financial time-series shows that our estimators capture essential aspects of market dynamics.\\ \\
\textbf{Keywords}:
random matrix theory, factor model, principal component analysis, free random variable, Kullback-Leibler divergence
\section{Introduction}
\label{sec::introduction}
%\subsection{Factor models in high-dimensions}
The increasing accessibility of `big data' occurs also in economics and finance. In dealing with such high-dimensional data sets, factor models are often used, since they can reduce the dimension and effectively extract relevant information. The estimation of high-dimensional factor models has been actively studied extensively in statistics and econometrics \cite{CR1983, ConnorKorajzcyk1986, ConnorKorajzcyk1993, SW1988, BaiNg2002, Kapetanios2004, Onatski2010, AH2013, Harding2013, BaiNg2013}. This paper provides a new approach to estimating high-dimensional factor models, using the eigenvalue distribution of residuals.
From a minimum distance approach, we estimate the number of factors and the correlation structure of residuals. The proposed method is validated with Monte carlo simulations and, in most of the cases we consider, it outperforms other known methods. Furthermore, the results with financial data show that our estimators effectively capture structural market changes.

Consider a factor model that is as follows. For $i=1,\cdots,N$ and $t=1,\cdots,T$,
\begin{eqnarray}
\label{eq::factormodel}
R_{it} = \sum_{j=1}^pL_{ij}F_{jt} + U_{it}
\end{eqnarray}
where $R_{it}$ is the data of $i$-th unit (e.g.,asset return) at time $t$, $F_{tj}$ is the $j$-th factor at time $t$, $L_{ij}$ is the loading of $j$-th factor on $i$-th cross-sectional unit. $U_{it}$ is the idiosyncratic component or residual of $R_{it}$. Usually, only $R$ is observable. Thus, the following questions are possible:
\begin{enumerate}
\item
How to estimate $F$ (factors) and $L$ (factor loadings)?
\item
How to estimate $p$ (number of factors)?
\item
$U=R-LF$. What are the properties of $U$ (residuals)? Are they noises or do they still contain information?
\end{enumerate}
For the first question, given $p$, principal components can be used to estimate $F$ and $L$. For the second and the third question, one way is to determine $p$ by looking at singular values of covariance matrix of $R$ and take some of them based on a given threshold for variance explanation. Then one usually assumes $U$ as pure noises.

However, in this paper, we mainly focus on the residuals $U$, and their dynamics and dependence, to estimate the covariance structures in $U$ and the number of factors $p$ simultaneously. Our approach is based on the investigation of the empirical spectral distribution of covariance matrix of residuals.%\footnote{In high dimensional settings: $N$ and $T$ go to infinity}.

%To estimate factors, principal components are used, not only because it is easy to do, but also because the method of asymptotic principal components consistently estimates the true factor space (\cite{ConnorKorajzcyk1986}; \cite{ConnorKorajzcyk1993}; \cite{Bai2003}; \cite{BaiNg2013}). However, the number of factors still needs to be estimated to use the principal components. Besides, in approximate factor models (\cite{CR1983}), residuals are assumed to have non-trivial correlation structure. Our method aims to simultaneously estimate the number of factors and parameters that determine the correlation matrices of residuals.

The first contribution of this paper is that we connect the factor model estimation problems to the limiting empirical eigenvalue distribution of covariance matrices of residuals. Thus, the main focus of the proposed method is on residuals, $U$. Instead of requiring that the idiosyncratic components $U_{it}$'s are uncorrelated to each other, we assume there are cross- and auto-correlated structures, such that $U$ is represented as $U=A_N^{1/2}\epsilon B_T^{1/2}$, where $\epsilon$ is an $N\times T$ $(T=T(N))$ matrix with i.i.d. Gaussian entries, and $A_N$ and $B_T$ are an $N \times N$ and $T \times T$ symmetric non-negative definite matrices, representing cross- and auto- covariances, respectively\footnote{This is not the most general model, since cross- and auto-covariance contributions are decoupled: $cov(U_{it}, U_{js}) = {A_N}_{ij}{B_T}_{ts}$.}. Then the empirical covariance matrix of $U$ can be written as
$C_N = \frac{1}{T}UU^T = \frac{1}{T}A_N^{1/2}\epsilon B_T\epsilon^TA_N^{1/2}$. In this paper, we restrict the matrix structures of $A_N$ and $B_T$, so that they are completely defined by simple parameter sets, $\theta_{A_N}$ and $\theta_{B_T}$ that are to be estimated along with the number of factors. For example, a simple case is that each residual has the same cross-correlation\footnote{We assume each time-series, $U_{it} \ \ (t=1, \cdots, T)$, is normalized and has an unit variance.}, $\beta$, to other residuals, and each residual has an exponentially decaying temporal auto–correlations with a parameter $\tau$. Then two parameters $\theta_{A_N} = \beta$ and $\theta_{B_T}=\tau$, completely determine $A_N$ and $B_T$, since $A_N = \big\{(A_N)_{ii} = 1, \ \ (A_N)_{ij, i\neq j} = \beta, \ \  i ,j=1,\cdots,N \big\}$ and $B_T =\big\{(B_T)_{st} = \exp(-|s-t|/\tau), \ \ s,t=1,\cdots,T\big\}$. %\footnote{As we will discuss later, we mainly focus on the temporal correlation structure of residuals, so in our simplified model, we present the result of the case where $A_N = I_{N\times N}$, supposing the cross-correlations are effectively removed by the factors.}

Now the objective of our estimation method is to match the eigenvalue distribution of $C_N$ to that of the empirical covariance matrix of residuals constructed from market data. The latter can be controlled by the number of principal components to be removed. The former depends on the modeling of $A_N$ and $B_T$, but we assume a parsimonious matrix structure, determined by only a small parameter set, ($\theta_{A_N}, \theta_{B_T}$).

We search for the number of factors ($p$) and the parameter sets ($\theta_{A_N}, \theta_{B_T}$), such that the spectral distance between a model and real data is minimized. This spectrum-based approach is motivated by the two typical characteristic aspects in the spectrum of real data: a few spikes and bulk. The former represent factors that mainly drive the market features and the latter arises from idiosyncratic noise. It is also theoretically motivated by the results of \cite{Zhang2006}, which analyzes, under certain assumptions, the convergence of the empirical eigenvalue distribution of $C_N$ to a suitable limiting distribution.

The factor model estimation problem is stated as follows.
\begin{eqnarray}
\label{eq::problem}
\{\hat p, \hat \theta\} = \arg\min\limits_{p, \theta} \mathcal{D}\Big(\rho_{\texttt{real}}(p), \rho_{\texttt{model}}(\theta)\Big)
\end{eqnarray}
where $\rho_{\texttt{real}}(p)$ is an empirical eigenvalue density of covariance matrix of residuals that are constructed by removing $p$ principal components from original data,  $\rho_{\texttt{model}}(\theta)$ is a limiting eigenvalue density of the general covariance matrix characterized by a parameter set $\theta = (\theta_{A_N}, \theta_{B_T})$, and $\mathcal{D}$ is a spectral distance measure or loss function we choose. The solution of this minimization problem gives the number of factors and parameters for the correlation structure of the residuals. As for estimating the number of factors, there are several methods proposed in previous literature \cite{BaiNg2002, Kapetanios2004, Onatski2010, AH2013, Harding2013}. The main difference from other estimators is that our method finds the best fit of the whole spectral distribution, which enables us to take into account both spikes and bulk of the distribution.

A difficulty is in the calculation of $\rho_{\texttt{model}}(\theta)$\footnote{$\rho_{\texttt{real}}(p)$ can be obtained easily with data. See Section \ref{sec::residualsfromdata} for details}, since using the limiting distribution from the Stieltjes transform in for general $A_N$ and $B_T$ is very complicated. However, a recent work by \cite{Burda2010} provides an analytic derivation of limiting spectral density using free random variable techniques. In this paper, we use these results to calculate $\rho_{\texttt{model}}(\cdot)$. Furthermore, we propose a simplified estimation problem that considers parsimonious matrix structures for $A_N$ and $B_T$. In particular, supposing that the cross-correlations are effectively removed by the factors, we assume that the cross-correlations among the normalized residuals are negligible: $A_N \thickapprox I_{N\times N}$ (or $\beta = 0$ in the previous example). But we still assume they are serially-correlated, with exponential decays with respect to time lags: $(B_T)_{ij} = b^{|i-j|}$. Then the $\rho_{\texttt{model}}(\theta_{A_N}, \theta_{B_T})$ is replaced by
$\rho_{\texttt{model}}(b)$, and the minimization problem has only two scalar variables, $p$ and $b$. This parsimonious model has significance in two senses. First, it is good for calculability, as we adopt the free-random variable techniques. Second, the parameter $b$ indicates global rate of mean-reversion of residuals. The mean-reversion property of residuals getting increasing attentions in the current financial markets, especially for statistical arbitrage strategy \cite{YP2017}.

The second main contribution of our work is that the proposed methods are validated from tests with synthetic data, generated using known models. Monte Carlo simulations with synthetic data show that the finite-sample performances of the estimators are good. The number of factors and the autoregressive parameter are accurately estimated for various choices for $N$ and $T$. We compare the estimated number of factors from our method with those from other methods in the literature, and show that our method is robust to noise and performs well in identifying weak factors.

The third contribution is that we find, with real market time-series data, that our estimators of the simplified problem successfully capture market dynamics. The estimation problem we propose is static, so in order to observe time-varying behaviors of parameters, we repeat the estimation procedures with moving windows. For market data, we use daily returns of S\&P500 stocks in the period of 2000-2015. We compute time changes of the estimators. It turns out that the estimators reflect the regime-change information of the market. In particular, we find that during stress periods, the number of factors is decreasing, while the variance explained by the corresponding factors increases, which shows market condensation. Furthermore, the \emph{global} mean-reversion time of residuals, represented by the estimated autoregressive coefficient $b$, tracks the volatility index very closely. We also find that during the crisis, the residuals are more trending, showing slower mean-reversions.

The rest of the paper consists of the following content. In Section \ref{sec::relatedliterature}, we review related literature. In Section \ref{sec::example1}, we consider a motivating example.
Section \ref{sec::factormodelestimations} describes our estimation method of factor models and describe the procedures used. Section \ref{sec::mc} contains Monte Carlo analysis and comparisons with other methods. Section \ref{sec::realdata} shows applications with real data. We conclude in Section \ref{sec::conclusions}.

\section{Related literature}
\label{sec::relatedliterature}
Our method in high-dimensional settings is fundamentally based on random matrix theory. Random matrix theory, developed originally to study the interactions in complex quantum systems \cite{Wigner1951}, can be used to identify non-random properties which are deviations from the universal predictions. \cite{LCBP1999} and \cite{RMT6} were the first two studies that applied the random matrix theory to financial correlations, and myriads of papers have followed in the physics community \cite{RMT4, RMT2, RMT1, DKMS2012, OEWJSK2011}. Comprehensive reviews on financial application of random matrix theory are available in \cite{BP2009} and \cite{BBP2017}. They have analyzed eigenvalue distribution of empirical cross-correlation matrix from stock returns. They claimed that deviated eigenvalues from a theoretical expectation, Marchenko-Pastur law \cite{MP1967}, provides genuine market information, such as market mode or industrial sectors. Then the number of factors is determined by counting those deviating eigenvalues.

However, ``no information'' or ``pure noise'' assumption in the bulk region\footnote{The eigenvalue distribution considered in this paper consists of many bounded small eigenvalues (bulk) and several large ones (spikes).} is too strict and it turns out to be invalid in practice. As seen from the example in Section \ref{sec::example1}, the fit of the empirical spectral density of covariance matrix from real residual returns to the Marchenko-Pastur distribution is problematic. This implies that the residuals from real data are not necessarily pure noise, and more general correlation structure needs to be considered to assess the empirical densities.

The phenomenal work by \cite{Zhang2006} provides a central theoretical foundation for our estimation method. The author considers a general covariance matrix, $C_N$, of the form $C_N = \frac{1}{T}A_N^{1/2}\epsilon B_T\epsilon^TA_N^{1/2}$, where $A_N$ and $B_T$ are non-negative definite matrices of size $N\times N$ and $T \times T$, respectively, and $\epsilon$ is an $N \times T$ Gaussian random matrix with i.i.d. entries. Let $c=N/T$. \cite{Zhang2006} shows that, under certain assumptions, the empirical eigenvalue distribution of $C_N$ converges weakly to a non-random distribution $\mathcal{F}^{{c,A,B}}$. In this paper, we introduce an approximate model with simple parameterizations, and directly derive the probability distribution of eigenvalues by using the techniques introduced in \cite{Burda2010}. Then we relate the spectrum of the model to real data.

%and that the Stieltjes transform of $\mathcal{F}^{{c,A,B}}$, $m(z)$, together with other analytical function $p(z)$ and $q(z)$, constitutes a solution to the system.\footnote{See Appendix for more detail}
%Assuming the residuals has the form $U=A_N^{1/2}\epsilon B_T^{1/2}$ in the factor model, this theory can be employed to show the consistency of our estimators from minimum spectral distance of residuals.

In the meantime, the factor model framework in finance was initiated by \cite{Ross1976} which proposed Arbitrage Pricing Theory. With relaxed assumptions allowing weak correlation in idiosyncratic components, approximate factor models were introduced by \cite{CR1983}. The dynamic factor models \cite{SW2005} also received attentions. Many physics researchers also have attempted to reveal correlation structures in financial market data using factor analysis \cite{GK2003, BLMP2007, Noh2000, MBGSM2005, LAS2011}.

The determination of the number of factors in high-dimensional factor models is one of the crucial issues in both theoretical and practical perspectives. The original work of \cite{BaiNg2002} uses an information criterion to determine the number factors. \cite{Kapetanios2004} is the first to use the idea of structure of idiosyncratic terms. The authoer points out that the correlated assumption on idiosyncratic components implies a closed-form expression for a sharp asymptotic upper bound on the idiosyncratic eigenvalues of the sample covariance matrix. Thus, he claims that counting the eigenvalues above the bound gives an estimate of the number of factors. \cite{Onatski2010} provides a criterion using the difference of two adjacent eigenvalues. The method based on the eigenvalues ratio is also developed in \cite{AH2013}, and recently in \cite{Pelger2016} for high-frequency data. \cite{Harding2013} also proposed a method for estimating the number of factors using spectrums. A difference from \cite{Harding2013} and ours is that the former takes only the first few moments, while our method uses the whole probability density, and takes into account the characteristic aspects of both spikes and bulk of the covariance matrix by using an appropriate metric. Thus, our method does not need to decide how many moments to take, and is free from the instability in using high-order moments. Furthermore, our study focuses on global mean-reversion rate, and investigates its dynamics with real data.
\section{Example: problematic fit of MP-law to real data}
\label{sec::example1}
In this section, we illustrate how much the Marchenko-Pastur (MP) \cite{MP1967} law can explain the spectrum of residuals after removing factors, from real market data and from synthetic data. As for real data, we obtain daily returns of 400 stocks in S\&P500 during 2012-2015 ($N=400$, $T=1000$):
\begin{eqnarray}
R^{real}_{it} = \frac{S_{it} - S_{i, t-1}}{S_{i, t-1}}.
\end{eqnarray}
where $S_{it}$ is the price of stock $i$ at time $t$.
Second, the synthetic data of the same dimension ($N=400$, $T=1000$) is generated by the following model %\footnote{This model is commonly used }
\begin{eqnarray}
R^{syn}_{it} = \sum\limits_{j=1}^p L_{ij}F_{jt} + U_{it}
\end{eqnarray}
where  $F_{jt} \sim N(0,0.1^2)$, $L_{ij}, U_{it} \sim N(0,1)$ are independent, and the true number of factors $p$ is set to be 3. That is, the correlation structure is known for synthetic data, while it is not the case not for real data.

Next, for each $R^{real}$ and $R^{syn}$, we construct \emph{p-level} residuals by removing factors, using principal components:
\begin{eqnarray}
\label{eq::plevelresidual}
{\hat U}^{(p)} = R - {\hat L}^{(p)} {\hat F}^{(p)}
\end{eqnarray}
where ${\hat L}^{(p)} {\hat F}^{(p)}$ is the estimated common factor from $p$ principal components. We are interested in the distribution of eigenvalues of covariance matrix of residuals ${\hat U}^{(p)}$:
\begin{eqnarray}
\label{eq::plevelcov}
{\hat C}^{(p)}  = \frac{1}{T} {\hat {U}}^{(p)}  {\hat {U}}^{(p)T}
\end{eqnarray}

\begin{figure}[!htbp]
\centering
\begin{tabular}{cc}
\includegraphics[width=0.3\linewidth]{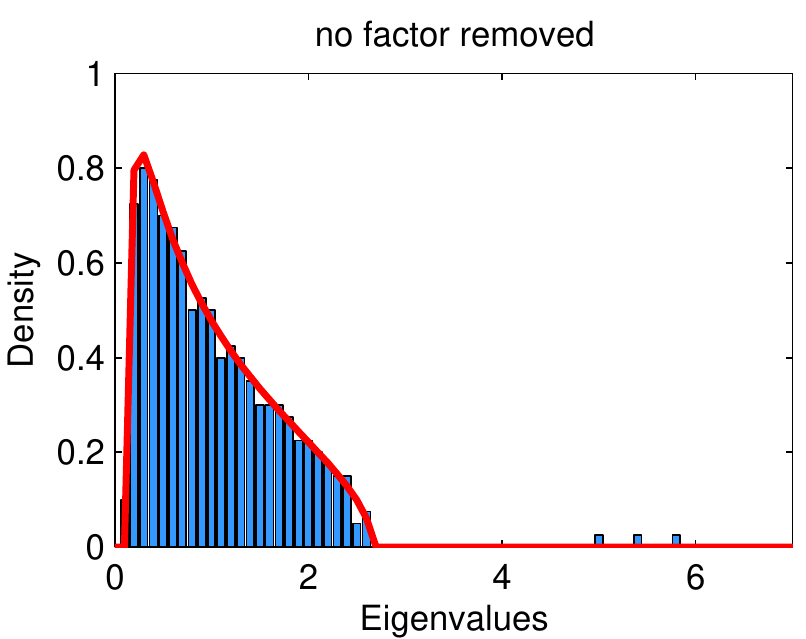}&
\includegraphics[width=0.3\linewidth]{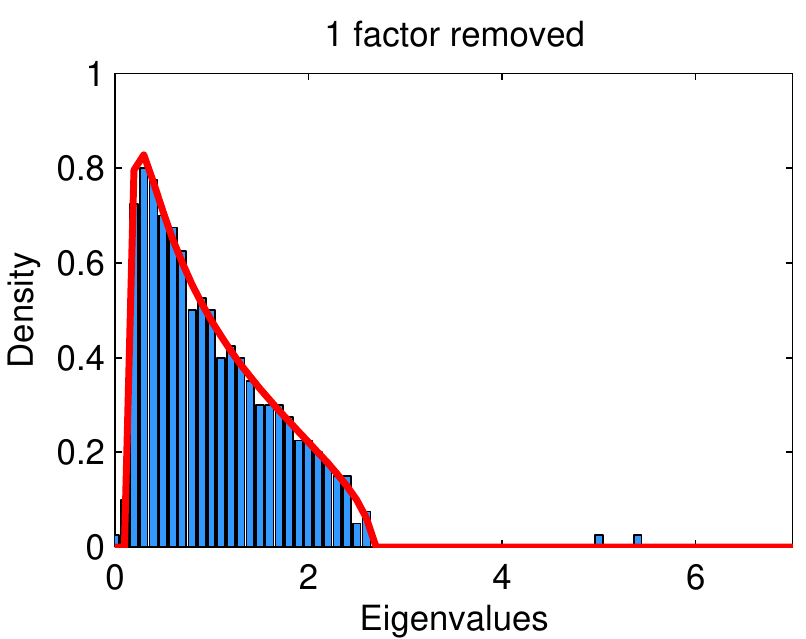}\\
\includegraphics[width=0.3\linewidth]{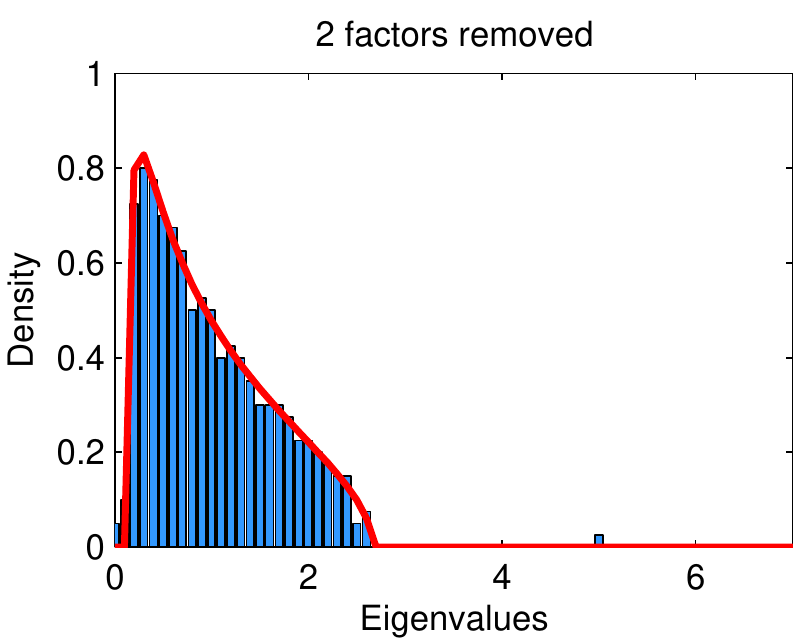}&
\includegraphics[width=0.3\linewidth]{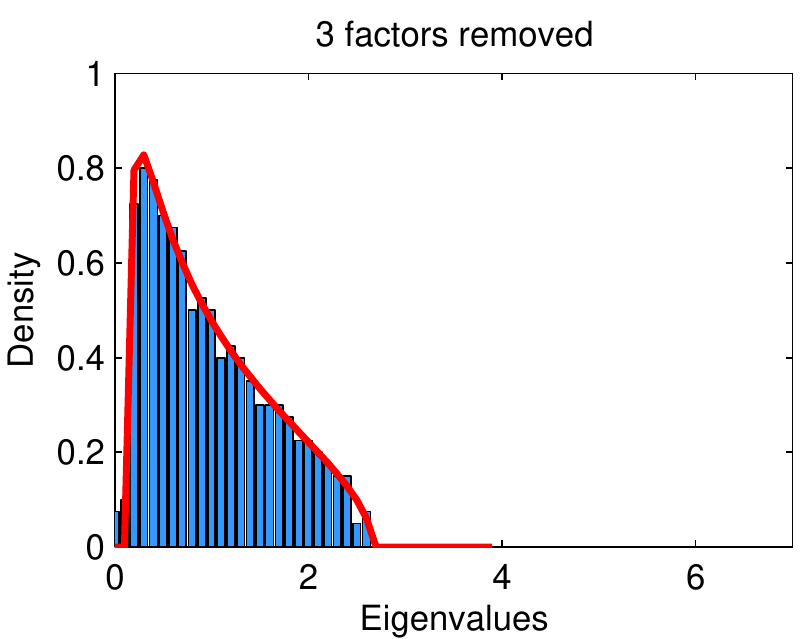}\\
\end{tabular}
\caption[Synthetic data]{Eigenvalue distribution of covariance matrix of residuals from synthetic data, when few principal components are removed. The true number of factors, $p$, is set to be 3. When 3 factors are removed, the corresponding spikes are all removed, and the remaining bulk part is well-fit by the Marchenko-Pastur (MP)-law.}
\label{fig::example1}
\end{figure}

\begin{figure}[!htbp]
\centering
\begin{tabular}{cc}
\includegraphics[width=0.3\linewidth]{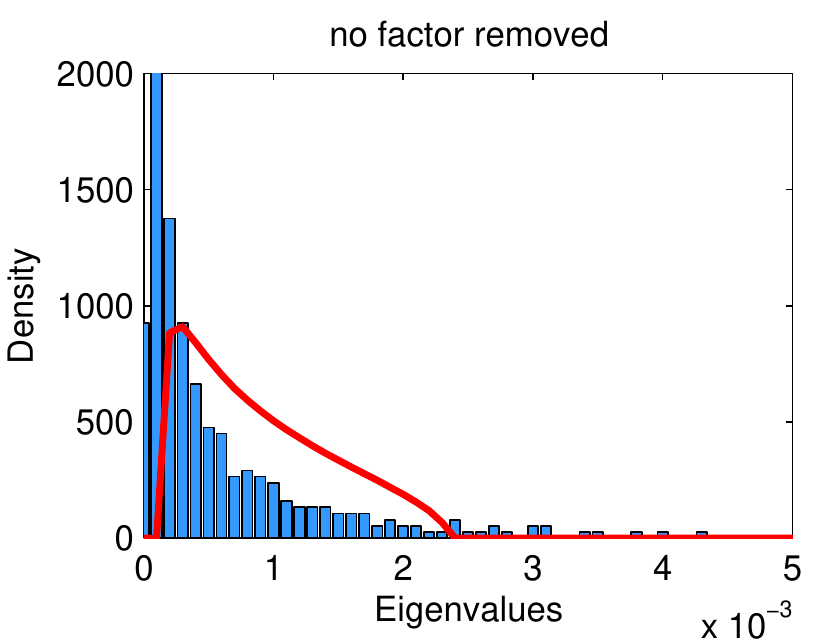}&
\includegraphics[width=0.3\linewidth]{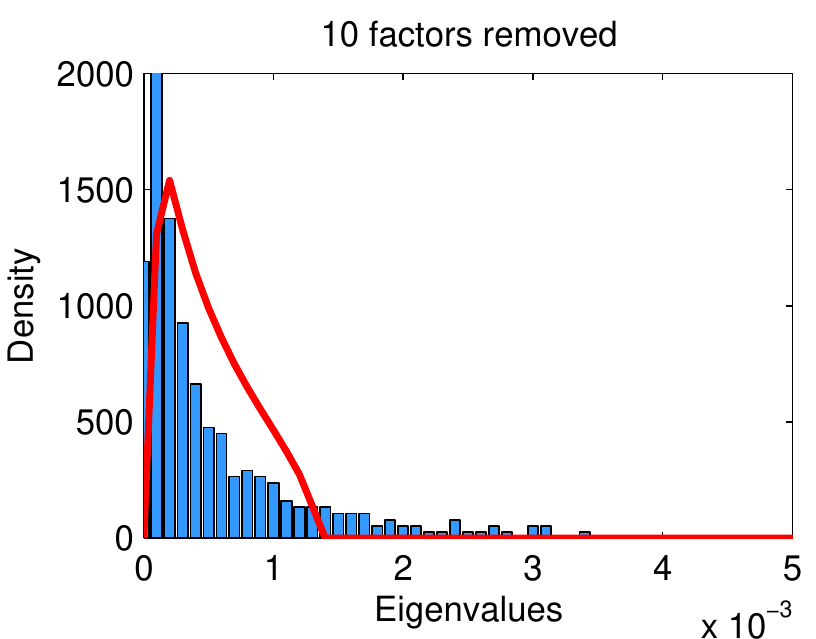}\\
\includegraphics[width=0.3\linewidth]{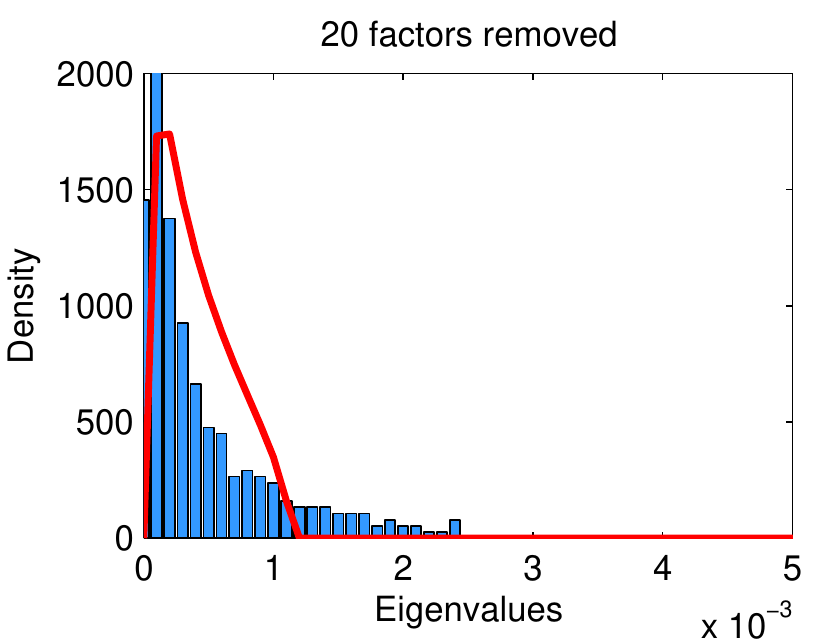}&
\includegraphics[width=0.3\linewidth]{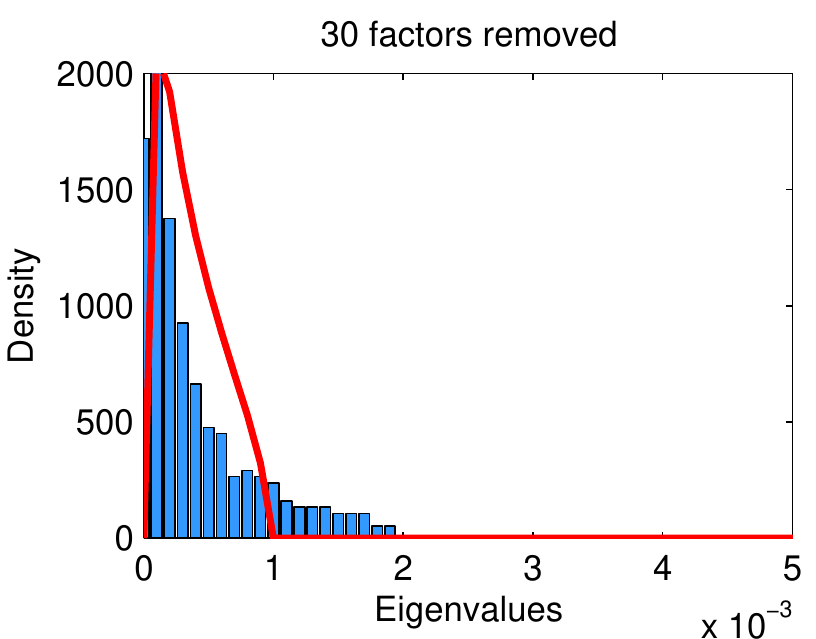}\\
\end{tabular}\\
\caption[Real Data]{Eigenvalue distribution of covariance matrix of residuals from real data. No matter how many factors are removed, the residual parts cannot be explained by MP-law. We also confirmed that using correlation matrix and its eigenvalues yields the same problem.}
\label{fig::example2}
\end{figure}

The eigenvalue distribution of residuals is depicted in Figure \ref{fig::example1}. As seen from the plot, the empirical spectrum consists of a bulk and few spikes. For the spectrum of raw data (no factor removed), there are three spikes, which corresponds to the three factors we generated. However, when the true number of factors (3) factors are removed, the spectral density of the residuals converges to the MP-law. On the contrary, as seen from Figure \ref{fig::example2}, the density with real data residual does not fit to the MP-law, no matter how many factors are subtracted. This experiment motivates us to develop the main idea of this paper: we allow correlations in $U$ and minimize spectral distance between the two distributions, to estimate factor models.

%confirmed that using correlation matrix and its eigenvalues has the same problem.

\section{Factor model estimations}
\label{sec::factormodelestimations}
\begin{figure}[!htbp]
\centering
  \includegraphics[width=12cm]{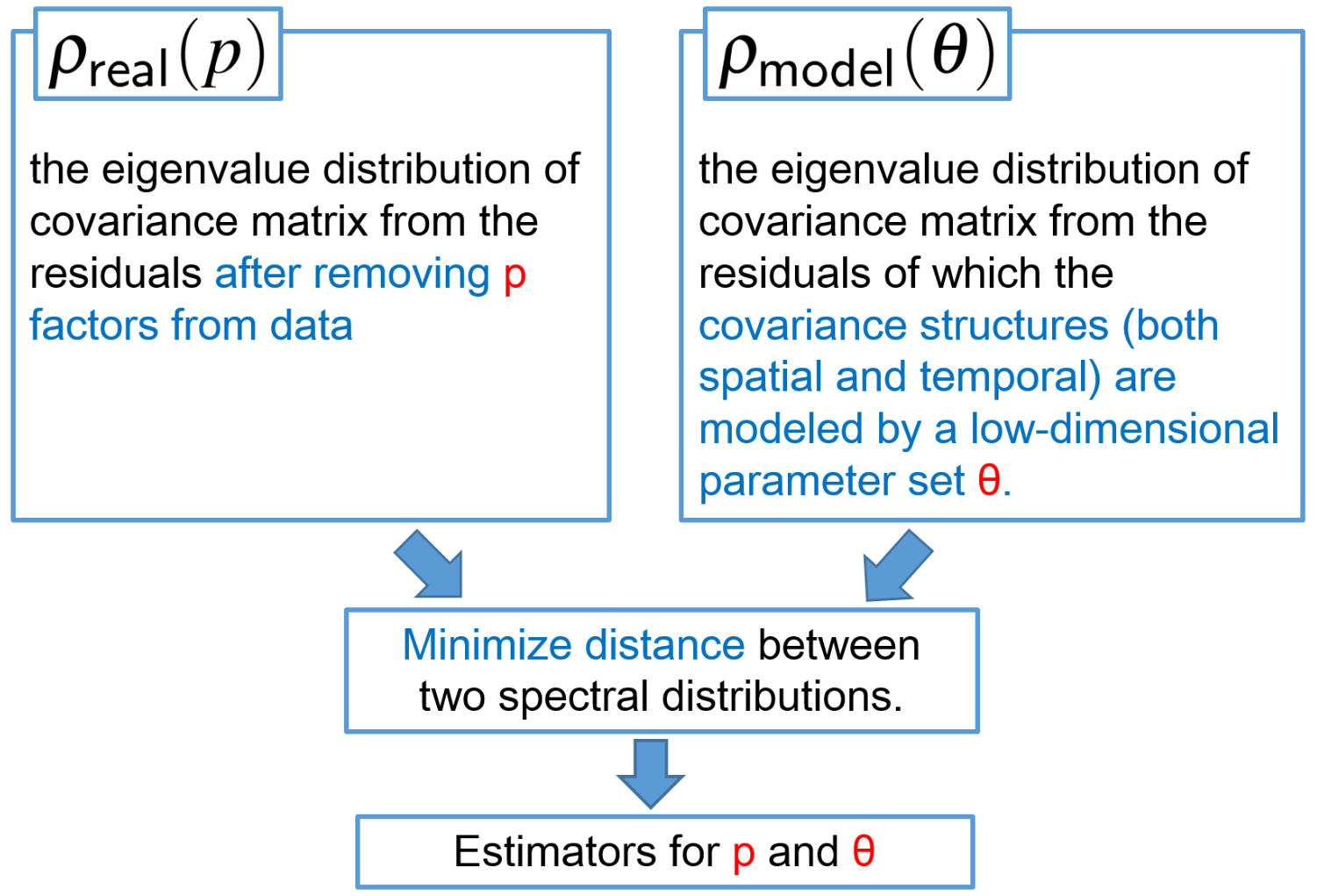}
\caption[Factor model estimation]{Schematic diagram for factor model estimation procedure. Based on minimum distance of spectra, it estimates the number of factors ($p$) and the parameter ($\theta$) for covariance structure of residuals.}
\label{fig::diagram_factormodelestimation}
\end{figure}

Our estimation method aims to find appropriate matches between two spectra. One is the empirical eigenvalue distribution of residuals that are obtained by removing factors from real data. The other is the empirical eigenvalue distribution of residuals of which the covariance structure is modeled by a parameter set. Once these two distributions are obtained, we minimize the distance between the two, so that we can estimate desired parameters. Our work is the first that applies this model to estimate covariance structures of residual returns from real data. Figure \ref{fig::diagram_factormodelestimation} illustrates the estimation procedures.
%Although Lemma \ref{lem::1} itself has nothing to do with factor models, it can be applied to the spectrum of covariance matrix of residuals in factor models. In this paper, we mainly discuss this connection, with high dimensional settings.

\subsection{$\rho_{\texttt{real}}$($p$): using principal components}
\label{sec::residualsfromdata}
The first step is to generate empirical residuals, by extracting $p$ largest principal components from real data. Here we use principal components as factors. In large dimensional data, principal components determine portfolios that approximately mimic all true factors up to rotations \cite{SW2002, Bai2003, FLM2013}. If more than one factor actually exists, p-level residual $\hat{U}^{(p)}$ in Eq. \ref{eq::plevelresidual}, can be always calculated for $p\geq1$. The covariance matrix from $p$-level residuals is given by Eq. \ref{eq::plevelcov}:
\begin{eqnarray}
C_{\texttt{real}}^{(p)} = \cfrac{1}{T}\hat{U}^{(p)}\hat{U}^{{(p)}T}.
\end{eqnarray}
The subscript $\texttt{real}$ indicates that it is constructed from real market data. We aim to find the number of factors from spectral distribution of $C_{\texttt{real}}^{(p)}$, by controlling $p$ in our algorithm. The idea behind this is simple. We keep subtracting factors until the bulk spectrum from the residuals using real data becomes close to that from modeled residuals.
\subsection{$\rho_{\texttt{model}}$($\theta$): modeling covariance of residuals}
The next step is to model the covariance structure of residual processes. Let the residuals have a certain covariance structure, characterized by parameters $\theta_{A_N}$ and $\theta_{B_T}$, for cross-covariance matrix $A_N$ and auto-covariance matrix $B_T$, respectively. Then we can suppose the residual term has a structure of the form\footnote{This model is known as Kronecker model, and widely used in communications \cite{KSPMF2002} and recently introduced in econometrics \cite{Onatski2010}.}
\begin{eqnarray}
U={A_N}^{1/2}\epsilon {B_T}^{1/2}
\end{eqnarray}
where $\epsilon$ is an $N \times T$ uncorrelated random matrix with i.i.d. entries, and $A_N$ and $B_T$ represent the cross- and auto- covariance structures, with parameter $\theta_A$ and $\theta_B$, respectively. Then the empirical covariance matrix of $U$ is given as
\begin{eqnarray}
C_N = \frac{1}{N}UU^T = \frac{1}{T}A_N^{1/2}\epsilon B_T \epsilon^T A_N^{1/2}
\end{eqnarray}
Note that if empirical spectral distribution of $A_N$ and $B_T$ converge, it is shown that the spectral distribution of $C_N$ converges to a suitable limit, when $N$ and $T$ are large (see Lemma \ref{lem::1} in Appendix)
\subsection{Spectral distance metric}
Since the empirical spectrum contains spikes, not all distance measures are useful in this problem. Our method needs a metric that must be sensitive to the presence of spikes as well as account for correctly reflect the distribution from grouped eigenvalues. We tested several distance metrics, for the covariance matrices we consider. We use
Jensen-Shannon divergence, which is a symmetrized version of Kullback-Leibler divergence.
\begin{eqnarray}
\mathcal{D}_{JS} (P\|Q) = \frac{1}{2}\mathcal{D}_{KL}(P\|M) +  \frac{1}{2}\mathcal{D}_{KL}(Q\|M)
\end{eqnarray}
where $P$ and $Q$ are probability densities, $M = \frac{1}{2}(P+Q)$ and  $\mathcal{D}_{KL}(P\|Q)$ is the Kullback-Leibler divergence defined by $\mathcal{D}_{KL}(P\|Q) = \sum\limits_i P_i \log \frac{P_i}{Q_i}$. Note that the Kullback-Leibler distance becomes larger if one density has a spike at a point while the other is almost zero at that point. Using this measure, in addition, the information disparity in the bulk region is also taken into account. Further discussion on its numerical calculation of Kullback-Leibler divergence with discretized grids is in Appendix \ref{app::sd}.

\subsection{Factor model estimation}
Now we are ready to state the estimation problem here. We solve a minimization problem which searches for an effective parameter set for covariance matrix of residual processes and the number of factors such that the distance between the spectrum from a model and that from real data is minimized.
\begin{eqnarray}
\label{eq::main}
\{\hat p, \hat \theta\} = \arg\min\limits_{p, \theta} \mathcal{D}\Big(\rho_{\texttt{real}}(p), \rho_{\texttt{model}}(\theta)\Big)
\end{eqnarray}
where $\rho_{\texttt{real}}(p)$ is the eigenvalue distribution of $C_{\texttt{real}}^{(p)}$, $\rho_{\texttt{model}}(\theta)$ is a limiting eigenvalue density of the general covariance matrix characterized by a parameter set $\theta = (\theta_{A_N}, \theta_{B_T})$, and $\mathcal{D}$ is a spectral distance measure or loss function we choose. This problem simultaneously estimates for the number of factors and parameters of residual correlations. The consistency of the estimators is discussed in Appendix \ref{consistency::1}.

\subsection{Simplified model on covariance structures of residuals}
As discussed earlier, the calculation of $\rho_{\texttt{real}}(p)$ is straightforward when using principal components estimators as factors. A difficulty lies in the calculation of the limiting distributions, $\rho_{\texttt{model}}(\theta)$, for general $\theta = (\theta_{A_N}, \theta_{B_T})$. Although Lemma \ref{lem::1} guarantees the convergence of empirical spectral distribution to a suitable limit, and the Stieltjes transforms obtained by the lemma provide useful information on the limiting distribution, the actual calculation of it is quite complex, which makes the implementation hard. However, a recent study of \cite{Burda2010} provides the direct derivation of spectral density using free random variable techniques. They particularly present analytic forms when the time-series follows vector autoregressive processes. In this paper, we employ this technique to calculate the spectrum $\rho_{\texttt{model}}(\cdot)$. For this, we propose a simplified modeling for $A_N$ and $B_T$, from mean-field model on spectrum of residual processes.

\subsubsection{Mean-field model on spectrum}
\label{sec::mft}
\begin{figure}[!htbp]
\centering
\begin{tabular}{c}
\includegraphics[width=0.8\linewidth]{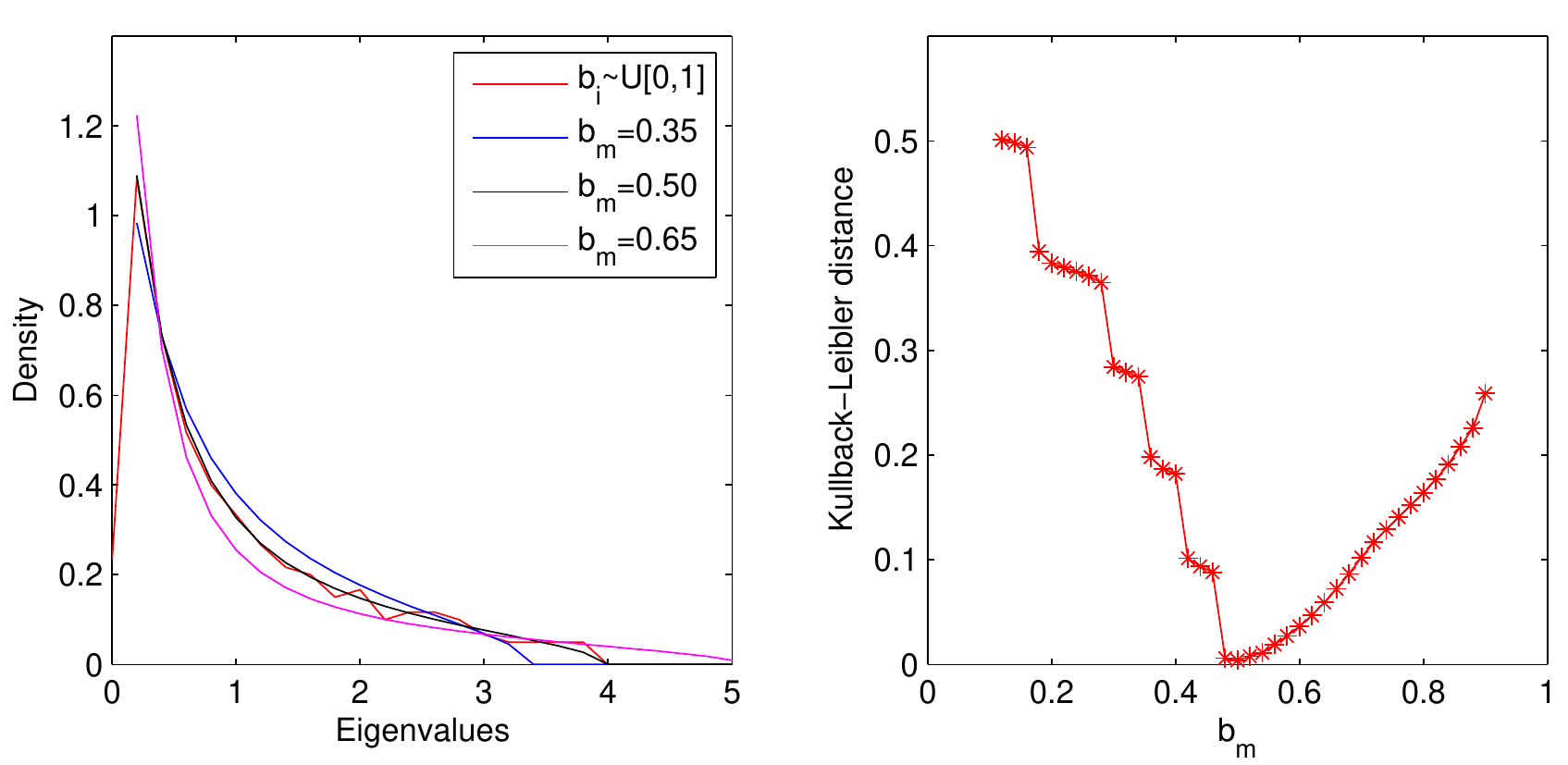}
\end{tabular}
\caption[Numerical demonstration of Claim \ref{prop::1}]{Numerical demonstration of Claim \ref{prop::1}. (Left): eigenvalue distribution from heterogenous process $Y$ with $b\sim U[0,1]$ (red), and from homogeneous autoregressive processes $Z$ with $\bar{b}$=0.35, 0.50, and 0.65, for $N=300, T=600$. (Right): Kullback-Leibler distance between $\rho_{C_Y}$ and $\rho_{C_Z}$. Note that the distance is minimized and almost zero near $\bar{b}=0.50$, which is actually the theoretical mean of $b_i$'s.}
\label{fig::meanfieldproof}
\end{figure}

A mean-field model is used to study the behavior of large and complex stochastic models by investigating a simpler model. For example, in magnetism in quantum spin systems, mean-field theory says that spin moves in the average field produced by all other spins. Usually in high dimensional systems, mean field theory gives a good picture of phase transitions. In factor models, each idiosyncratic return has its own driving force, namely a field. Analogous to traditional mean-field theory, rather than considering every individual residual separately, we consider single correlation structure that enables us to approximately replicate the spectral density of the original heterogenous correlation structures.
\begin{claim}[Mean-field model on spectrum]\label{prop::1}
Suppose we have two $N \times T$ matrices, $Y$ and $Z$, such that
\begin{eqnarray}
\label{eq::meanfield}
Y_{it} = b_iY_{i,t-1} + \xi_{it}\\
Z_{it} = \overline{b}Z_{i,t-1} + \eta_{it}
\end{eqnarray}
where $|b_i| < 1$, $\overline{b} = \frac{1}{N}\sum\limits{b_i}$, $\xi_{it}\sim N(0,\sigma_i^2)$ and $\eta_{it}\sim N(0,\overline{\sigma}^2)$. Let $\sigma_i^2 = 1-b_i^2$ and $\overline{\sigma}^2 = 1- \overline{b}^2$, so that $\var(Y)= \var(Z)=1$. Consider two empirical spectral distributions, $\rho_{C_Y}$ and $\rho_{C_Z}$, where
$C_Y = \frac{1}{T}YY^T$ and $C_Z = \frac{1}{T}ZZ^T$. Then the distance between $\rho_{C_Y}$ and $\rho_{C_Z}$ becomes sufficiently small, as $N,T$ are large.
\begin{eqnarray}
\mathcal{D}\Big(\rho_{C_Y}, \rho_{C_Z}\Big)\thickapprox 0
\end{eqnarray}
\end{claim}
For this claim, we provide a numerical illustration. We first draw random numbers for $b_i$, from a uniform distribution between 0 and 1, and take several different $\bar{b}$ values, $\bar b = 0.35, 0.5, 0.65$. The synthetic data sets for $Y$ and $Z$ are generated from the above autoregressive processes in Eq. \label{eq::meanfield}. In Figure \ref{fig::meanfieldproof}, we present the eigenvalue distribution $C_Y$ and $C_Z$. Among the cases of $\bar b = 0.35, 0.5, 0.65$, we discovered that the spectrum of $\rho_{C_Y}$ (red line) is the closest to $\rho_{C_Z}$ when $\bar {b}=0.5$ (black line), and the spectral distance (Kullback-Leibler distance in this case) is minimized at the same point.
\subsubsection{Factor model estimation with simplified model}
\label{sec::simpleparameters}
Now we propose a modified model, which has much simpler parameter sets, for $A_N$ and $B_T$. Suppose the following:
\begin{enumerate}
\item
The cross-correlations are effectively removed from $p$ principal components, where $p$ is the true number of factors, and the residual $U^{(p)}$ has sufficiently negligible cross-correlation: $A_N \thickapprox I_{N\times N}$.
\item
The autocorrelations of $U$ are exponentially decreasing (by an identical rate) with respect to time-lags: $\big\{{B_T}\big\}_{ij} = b^{|i-j|}$, with $|b|<1$. (This is equivalent to modeling residual returns as an AR(1) process: $U_{it} = b U_{i,t-1} + \xi_{it}$, where $\xi_{it}\sim N(0,1-b^2)$ so that the variance of $U_t$ is one.)
\end{enumerate}
From these assumptions and the mean-field model on spectrum in the previous section, we approximate the original estimation by using only two control variables, the number of factors, $p$, and the global mean-reversion rate $b$. In short, the estimation with simplified parameterizations is stated as
\begin{eqnarray}\label{eq::mod}
\{\hat{p},\hat {b}\} = \arg\min\limits_{p, b} \mathcal{D}\Big(\rho_{\texttt{real}}(p), \rho_{\texttt{model}}(b)\Big).
\end{eqnarray}
For numerical experiments in the following sections, we work with this simplified model. Although it seems to be too simple at the first glance, we will show that it sufficiently improves the robustness to noise levels and the ability of detecting weak factors.
%\begin{consistency}
%\label{consistency::2}
%Suppose the Assumptions \ref{as::1} to \ref{as::6} hold. If the number of factors is $p^*$ and the $U_{it}$ follows vector AR(1) processes with identical coefficient $b^*$ for all $i=1, \cdots, N$, then the solution of Eq. \ref{eq::mod}, $\{\hat{p},\hat {b}\}$ converges to  $\{p^*, b^*\}$, as $N,T\rightarrow \infty$.
%\end{consistency}

\subsubsection{Calculation of $\rho_{\texttt{model}}(b)$}
The simplified problem enables us to calculate the modeled spectral density, $\rho_{\texttt{model}}(b)$, more easily. It can be done by using the free random variable techniques proposed in \cite{Burda2010}. We briefly describe the major implementations here.
\begin{enumerate}
\item
The mean spectral density can be derived from the Green's function $G(z)$ by using the Sokhotsky's formula:
\begin{eqnarray}
\rho_{\texttt{model}}(\lambda) = -\frac{1}{\pi}\lim\limits_{\epsilon \rightarrow 0^+} \mathfrak{Im} G_\mathbf{c} (\lambda + i\epsilon).
\end{eqnarray}
\item
The green's function $G(z)$ can be obtained from the moments' generating function $M(z)$.
\begin{eqnarray}
M(z) = zG(z) - 1
\end{eqnarray}
\item
M(z) can be found by solving the polynomial equation for $M = M(z)$ ($a = \sqrt{1-b^2}$ and $c = N / T$):
\begin{eqnarray}
\label{eq::poly}
a^4c^2M^4 + 2a^2c\big(-(1+b^2)z+a^2c\big)M^3 &+&\\
\big((1-b^2)^2z^2-2a^2c(1+b^2)z + (c^2-1)a^4\big)M^2-2a^4M-a^4&=&0\nonumber
\end{eqnarray}
\end{enumerate}
See Appendix \ref{sec::frv} for details.
%\subsection*{Remark}
%Since we are using statistical factor models, the only known quantity is the time-series of returns, $R$. In order to systemically investigate multivariate time-series via the factor model, one needs to identify: (1) the number of factors, $p$, (2) factors $F$ and their loadings $B$, and (3) modeling idiosyncratic parts, $U$. Since these are interconnected to each other, we take into account their relations in the estimation program.
\subsubsection*{Remarks}
Although this simplified model came from our assumptions on covariance matrices, it actually has several benefits. First, it makes the calculation of the density almost analytically. The numerical process to obtain the spectral density $\rho_{\texttt{model}}(b)$ is straightforward, if we use the free-random variable techniques. Second, the two parameters reflect the essential features of typical spectra of covariance matrices we considers. As shown before, the spectrum is roughly decomposed into two parts: spikes and a bulk. The parameter $p$ controls the number of spikes in the residuals. As we subtract $p$ factors from data, then $p$ spikes that correspond to the $p$ largest eigenspaces are removed from the spectrum of the original data. At the same time, the parameter $b$ controls the region of smaller eigenvalues. Although it does not represent all possible shapes of bulks, it can effectively emulate the variability of the bulk spectrum of residuals. Based on the numerical results, it turns out that the edge of the bulk is sufficiently controllable within the desired numerical precisions. In addition, we also found from the Monte Carlo simulations that the number of factors is still accurately estimated by the method that uses only $b$. Third, the parameter $b$ is an aggregate quantity that represents the rate of mean-reversion of residual returns. The dynamics of residual spaces has received a significant attention in recent years. Although it cannot directly be applied to any practical use such as trading, the characterization of residual subspace of real markets using this single parameter provides an insight into market dynamics.%\footnote{A possible outcome of the simplified model is that, if the true residual processes are only cross-correlated but not auto-correlated, then the solution of the minimization problem gives a number for $b$, which may be far different from the true value. From a spectrum point of view, this number is equivalent to the true number.}

\section{Monte Carlo analysis}
\label{sec::mc}
\subsection{Experiments setup}
We evaluate the performance of our estimation method by Monte Carlo studies. We first generate synthetic data, using the following model:
\begin{eqnarray}
\label{eq::synthetic}
X_{it} &=& \sum\limits_{j=1}^p L_{ij}F_{jt} + \sqrt{\theta}U_{it};\\
with \ \ && \\
U_{it} &=& \sqrt{\cfrac{1-\rho^2}{1+2J\beta^2}} e_{it} \\
where \ \ && \\
e_{it} &=& \rho e_{i,t-1} + v_{it} + \sum\limits_{h=\max(i-J, 1)}^{i-1}\beta v_{ht} + \sum\limits_{h=i+1}^{\min(i+J, N)}\beta v_{ht} \\
v_{ht}, L_{it}, F_{jt} &\sim& N(0,1)
\end{eqnarray}
This model is also used in other papers \cite{BaiNg2002, Onatski2010, AH2013}. The rationale of this model is as follows.
\begin{enumerate}
\item
The coefficient $\sqrt{\cfrac{1-\rho^2}{1+2J\beta^2}}$ makes the variance of $U_{it}$ be always 1. This allows the model to control the variance (or noise) level of residuals only by $\theta$.
\item
$\theta$ controls the \emph{signal-to-noise ratio} (SNR), where $SNR = \frac{\var(Factors)}{\var(Residuals)} = \frac {p}{\theta}$. We will use 1/SNR instead of $\theta$ to denote the noise level. For example, if 1/SNR = 0.25, this implies $\theta=0.25 \times p$.
\item
$\rho$ controls the decaying rate of auto-correlations of residuals. ($|\rho| < 1$)
\item
Cross-correlations of residuals are controlled by $\beta$ for magnitudes $|\beta| \leq 1$ and by $J$ for affecting ranges. Since this local cross-correlations can be broader for larger system in practice, we set $J$ is proportional to $N$, i.e., $J=N/10$.
\end{enumerate}

The model parameters used in our Monte Carlo analysis are summarized in Table \ref{table::syntheticparameters}.%\footnote{We also have tried other possible values, but there is no other significant implications found from those.}
\begin{table}[htp]
\center
\scalebox{0.90}{
    \centering
    \begin{tabular}{|c|c|c|}
        \hline
        Sample sizes & $N,T$ &   $\{50,100,200,300,500\}$ \\ \hline
        Number of factors  & $p$ & $\{3,4,5\}$ \\ \hline
        1/SNR & $\theta$ & $\{0.1,0.25,0.5,0.75,1,1.5,2,3\}\times p$ \\ \hline
        Correlations in residuals & ($\rho,\beta,J$)  & $\{(0, 0, 0), (0.5, 0, 0), (0, 0.5, N/10), (0.5, 0.5, N/10) \}$ \\
        \hline
    \end{tabular}}
    \caption[Parameter configurations used in the Monte carlo experiments.]{Parameter configurations used in the Monte carlo experiments. }
    \label{table::syntheticparameters}
\end{table}

We first investigate the performance of our method, by checking the estimated values with true ones. Next, we focus on the number of factors. The estimated number of factors from our method is compared with those came from other three methods of \cite{BaiNg2002}, \cite{Onatski2010}, and \cite{AH2013}. For this, we examine several perspectives: (1) the convergence rate of error when the sample size becomes small or large, (2) the effect of the different residual correlation structures on the estimation error, and (3) the performance with various noise levels. Lastly, we tested the detection ability in the presence of weak factors.

As an error measure, the root mean squared error (RMSE) is obtained over 1000 replications. Before computing eigenvalues and eigenvectors, each series is demeaned and standardized to have unit variance.

%\begin{claim}\label{claim::2}
%Suppose the assumptions in Corrolary \ref{thm::2} hold. Let $p_{\texttt{orig}}$ and $p_{\texttt{mod}}$ are the estimator for the number of factors from the orignal problem (Eq.\ref{eq::problem}) and the modified problem (Eq.\ref{eq::mod2}):
%\begin{eqnarray}
%\{\hat p_{\texttt{orig}}, \hat \theta_{A_N}, \hat \theta_{B_T}\} &=& \arg\min\limits_{p, \theta_{A_N}, \theta_{B_T}} \mathcal{D}\Big(\rho_{\texttt{real}}(p), \rho_{\texttt{model}}(\theta_{A_N}, \theta_{B_T};c)\Big )\\
%\{\hat p_{\texttt{mod}},\hat {b}\} &=& \arg\min\limits_{p, b} \mathcal{D}\Big(\rho_{\texttt{real}}(p), \rho_{\texttt{model}}(b; c)\Big).
%\end{eqnarray}
%Then $p_{\texttt{orig}} \approx p_{\texttt{mod}}$.
%\end{claim}

\subsection{Estimation performance}
We first check the performance of our method on estimating factor models. Table \ref{table::MC_yeo} summarizes the averages of $\hat p$ and $\hat b$. We can first observe that the averages of our estimators, $\hat p$ and $\hat b$, are very close to the true number of factors and true auto-correlation coefficient for a broad range of $N$ and noise 1/SNR. One exception is when the sample size is small and the noise amount is large, where  our estimator starts to underestimate the true number of factors.

The true correlation structures are also varied in the test. $\rho$ represents the identical auto-regressive coefficient for residuals and $\beta$ represents the cross-correlation within the range of $J$ in the matrix. For the first case where there is no correlation in residuals, as $(\rho,\beta)=(0,0)$, the estimator $\hat b$ gives numbers between 0.03 and 0.05 which is close to the true value 0. When auto-correlations are imposed, as $(\rho,\beta)=(0.5 ,0)$, $\hat b$ is also very close to the true value 0.5. Adding cross-correlation structure here, as $(\rho,\beta)=(0.5 ,0.5)$, shifts the average value and decreases the accuracy, but not significantly. This is due to the fact that in our experiment setup the contribution of local cross-correlations on the spectrum is insignificant to that of auto-correlations. However, when only cross-correlations are forced in true residual processes, as $(\rho,\beta)=(0 ,0.5)$, the average of $\hat b$ is going far from the true value 0, giving values between 0.1 and 0.25. We interpret that from a spectral point of view, this deviated $\hat b$ has an effect on the spectrum approximately \footnote{within numerical tolerance on the spectral distance} equivalent the contribution from cross-correlations. However, we emphasize that this cross-correlation-only structure does not decrease the accuracy $\hat p$, as seen from the table.
\begin{table}[!htbp]
\centering
\scalebox{0.85}{
    \begin{tabular}{|c|c||c|c||c|c||c|c||c|c|}
        \hline
  \multicolumn{2}{|c||}{} &
  \multicolumn{2}{c||}{$\rho,\beta=(0,0)$} &
  \multicolumn{2}{c||}{$\rho,\beta=(0.5,0)$} &
  \multicolumn{2}{c||}{$\rho,\beta=(0.5,0.5)$} &
  \multicolumn{2}{c|}{$\rho,\beta=(0,0.5)$} \\
  \hline
        $N,T$ & 1/SNR & $\hat{p}$ & $\hat{b}$ & $\hat{p}$ & $\hat{b}$ & $\hat{p}$ & $\hat{b}$ & $\hat{p}$ & $\hat{b}$\\
\hline								
50	&	0.10	&	4.000	&	0.048	&	4.006	&	0.489	&	4.006	&	0.495	&	4.000	&	0.217	\\	
50	&	0.25	&	4.000	&	0.047	&	4.024	&	0.483	&	4.031	&	0.489	&	4.000	&	0.200	\\	
50	&	0.50	&	4.011	&	0.047	&	4.041	&	0.478	&	4.052	&	0.484	&	4.002	&	0.194	\\	
50	&	0.75	&	4.034	&	0.047	&	3.999	&	0.478	&	4.002	&	0.484	&	4.002	&	0.192	\\	
50	&	1.00	&	4.069	&	0.047	&	3.817	&	0.489	&	3.826	&	0.493	&	3.987	&	0.192	\\	
50	&	1.50	&	4.086	&	0.047	&	3.657	&	0.495	&	3.564	&	0.502	&	3.822	&	0.211	\\	
50	&	2.00	&	4.030	&	0.046	&	3.604	&	0.485	&	3.616	&	0.490	&	3.541	&	0.242	\\	
50	&	3.00	&	3.665	&	0.045	&	3.653	&	0.456	&	3.560	&	0.465	&	3.434	&	0.251	\\	\hline
100	&	0.10	&	4.000	&	0.050	&	4.000	&	0.504	&	4.000	&	0.505	&	4.000	&	0.172	\\	
100	&	0.25	&	4.000	&	0.049	&	4.000	&	0.503	&	4.000	&	0.503	&	4.000	&	0.162	\\	
100	&	0.50	&	4.000	&	0.049	&	4.000	&	0.502	&	4.000	&	0.502	&	4.000	&	0.159	\\	
100	&	0.75	&	4.000	&	0.049	&	4.000	&	0.501	&	4.001	&	0.501	&	4.000	&	0.158	\\	
100	&	1.00	&	4.000	&	0.049	&	4.000	&	0.501	&	4.001	&	0.501	&	4.000	&	0.159	\\	
100	&	1.50	&	4.002	&	0.049	&	4.001	&	0.500	&	4.001	&	0.500	&	4.000	&	0.158	\\	
100	&	2.00	&	4.002	&	0.049	&	3.973	&	0.501	&	3.991	&	0.500	&	4.000	&	0.157	\\	
100	&	3.00	&	4.008	&	0.049	&	3.640	&	0.517	&	3.813	&	0.511	&	3.992	&	0.158	\\	\hline
150	&	0.10	&	4.000	&	0.039	&	4.000	&	0.505	&	4.000	&	0.505	&	4.000	&	0.132	\\	
150	&	0.25	&	4.000	&	0.038	&	4.004	&	0.505	&	4.004	&	0.504	&	4.000	&	0.119	\\	
150	&	0.50	&	4.000	&	0.038	&	4.015	&	0.505	&	4.009	&	0.504	&	4.000	&	0.113	\\	
150	&	0.75	&	4.000	&	0.038	&	4.017	&	0.505	&	4.015	&	0.504	&	4.000	&	0.112	\\	
150	&	1.00	&	4.017	&	0.038	&	4.019	&	0.505	&	4.020	&	0.504	&	4.001	&	0.111	\\	
150	&	1.50	&	4.061	&	0.039	&	4.038	&	0.504	&	4.037	&	0.503	&	4.001	&	0.110	\\	
150	&	2.00	&	4.065	&	0.039	&	4.060	&	0.503	&	4.061	&	0.502	&	4.005	&	0.109	\\	
150	&	3.00	&	4.060	&	0.039	&	4.114	&	0.501	&	4.099	&	0.501	&	4.017	&	0.106	\\	\hline
200	&	0.10	&	4.000	&	0.050	&	4.000	&	0.506	&	4.000	&	0.507	&	4.000	&	0.118	\\	
200	&	0.25	&	4.000	&	0.050	&	4.000	&	0.506	&	4.000	&	0.506	&	4.000	&	0.114	\\	
200	&	0.50	&	4.000	&	0.050	&	4.000	&	0.505	&	4.000	&	0.506	&	4.000	&	0.112	\\	
200	&	0.75	&	4.000	&	0.050	&	4.000	&	0.505	&	4.000	&	0.506	&	4.000	&	0.112	\\	
200	&	1.00	&	4.000	&	0.050	&	4.002	&	0.505	&	4.001	&	0.506	&	4.000	&	0.112	\\	
200	&	1.50	&	4.000	&	0.050	&	4.002	&	0.505	&	4.002	&	0.506	&	4.000	&	0.112	\\	
200	&	2.00	&	4.005	&	0.050	&	4.001	&	0.505	&	4.002	&	0.505	&	4.000	&	0.112	\\	
200	&	3.00	&	4.061	&	0.050	&	4.008	&	0.505	&	4.004	&	0.505	&	4.000	&	0.111	\\	

        \hline
    \end{tabular}}
    \caption[Average values of the estimated $p$ and $b$ over 1000 simulations]{Average values of the estimated $p$ and $b$ over 1000 simulations. There are four different residual correlation structures: $(\rho, \beta) = (0, 0), (0.5, 0), (0.5, 0.5), (0, 0.5)$, and $J=N/10$. True number of factors is $p=4$. Note that if $\beta=0$, $\hat{b} $ must be an estimator of $\rho$, since in this case, the generating model for synthetic data and our assumed model for reduced problem are exactly the same. Otherwise, $\hat{b}$ does not necessarily converge to $\rho$, as seen from the last column, for example. The tables for RMSE for each estimate is provided in a supplemental report.}
    \label{table::MC_yeo}
\end{table}

\subsection{Comparison with other methods}
In this section, we will compare estimators from our spectral distance (SD, hereafter) method with other methods, including
the BIC3 estimator of \cite{BaiNg2002} that uses information criteria, the ED estimator of \cite{Onatski2010} that uses eigenvalue differences, and the ER estimator of \cite{AH2013} that uses eigenvalue ratios.
\subsubsection{Sample sizes and noise amounts}
\begin{figure}[!htbp]
\centering
\scalebox{0.85}{
\begin{tabular}{| >{\centering\arraybackslash}m{1in} | >{\centering\arraybackslash}m{2.4in}  >{\centering\arraybackslash}m{2.4in}  |}
\hline
 & \texttt{cross-correlations} (0, 0.5) & \texttt{auto- and cross-correlations} (0.5, 0.5)\\
\hline
1/SNR=0.25 &
\includegraphics[width=0.8\linewidth]{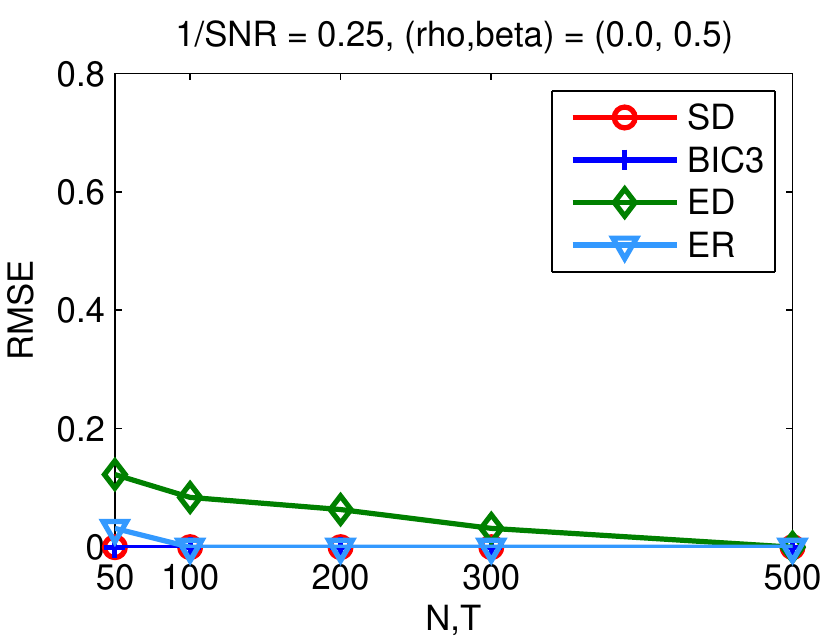}&
\includegraphics[width=0.8\linewidth]{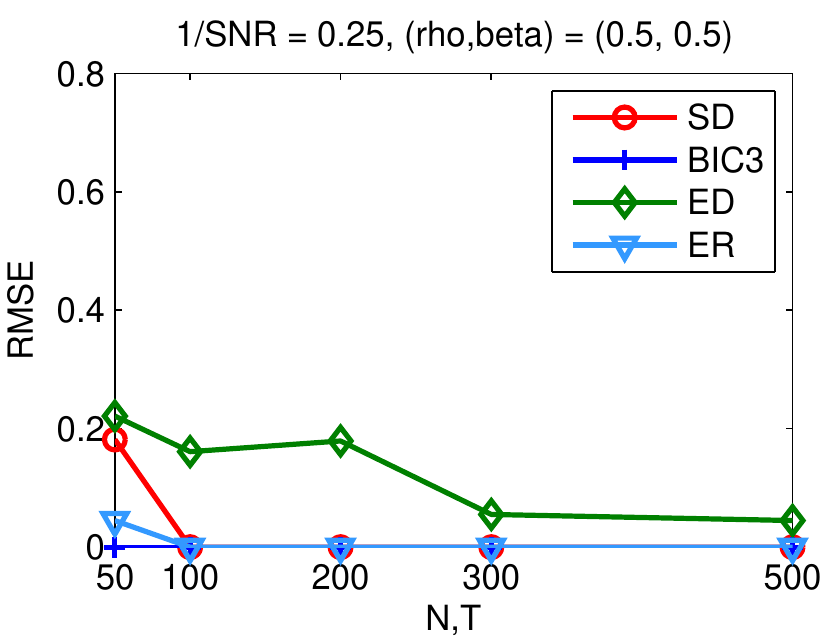}\\
1/SNR=0.75 &
\includegraphics[width=0.8\linewidth]{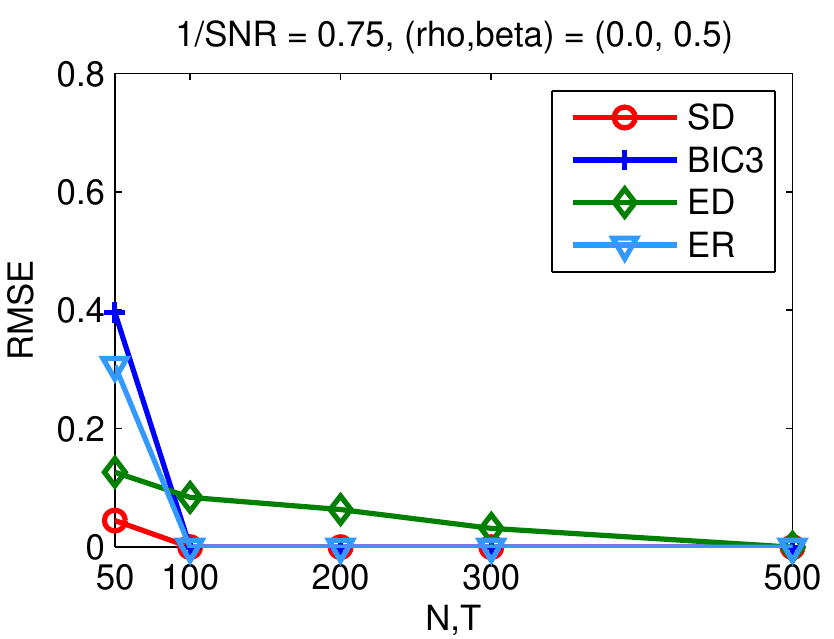}&
\includegraphics[width=0.8\linewidth]{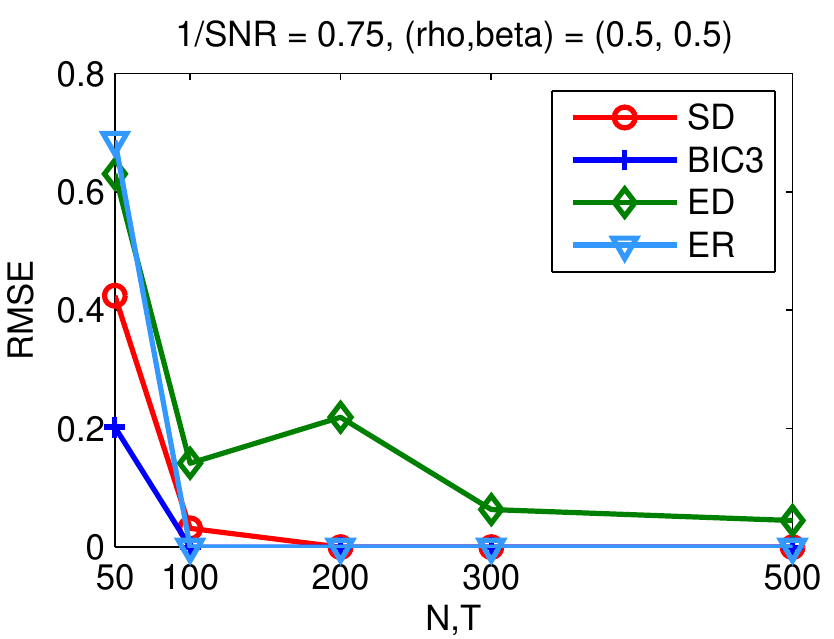}\\
1/SNR=1.00  &
\includegraphics[width=0.8\linewidth]{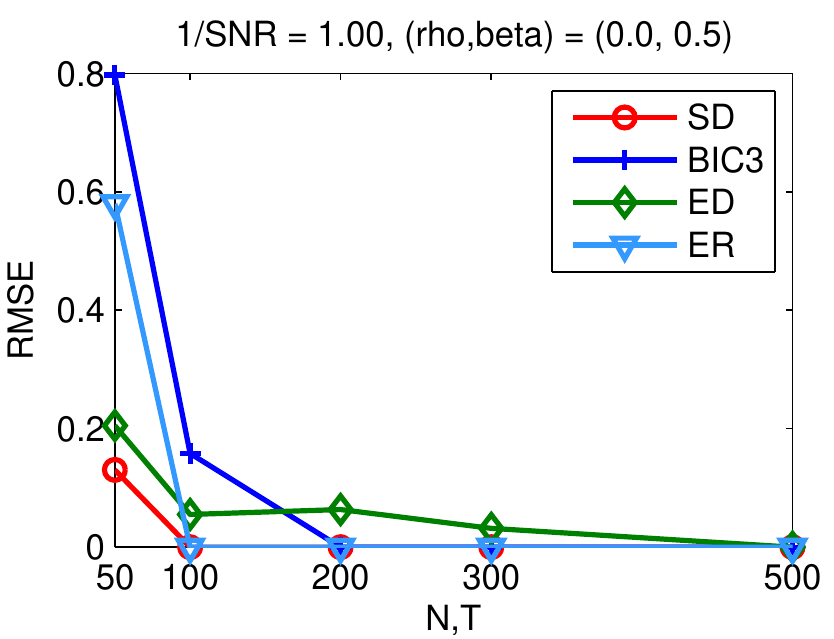}&
\includegraphics[width=0.8\linewidth]{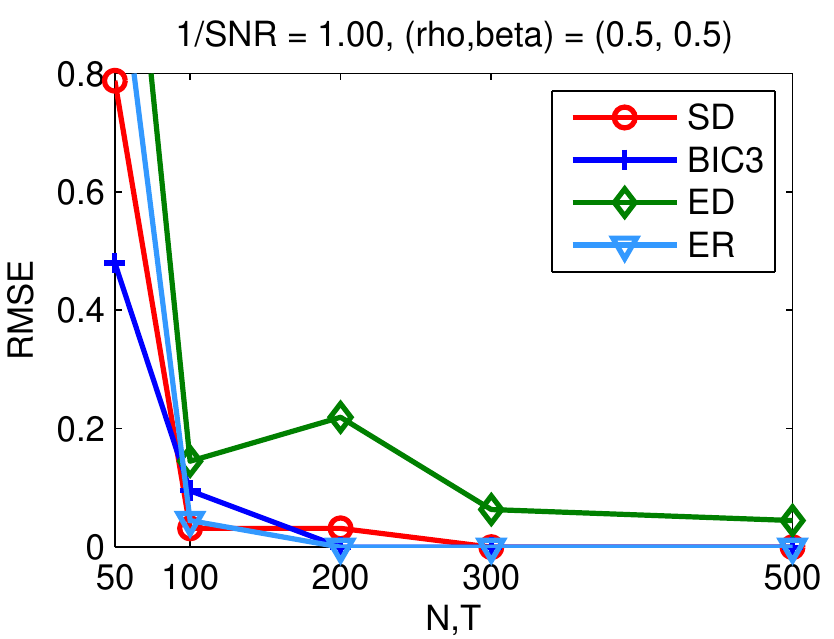}\\
1/SNR=2.00  &
\includegraphics[width=0.8\linewidth]{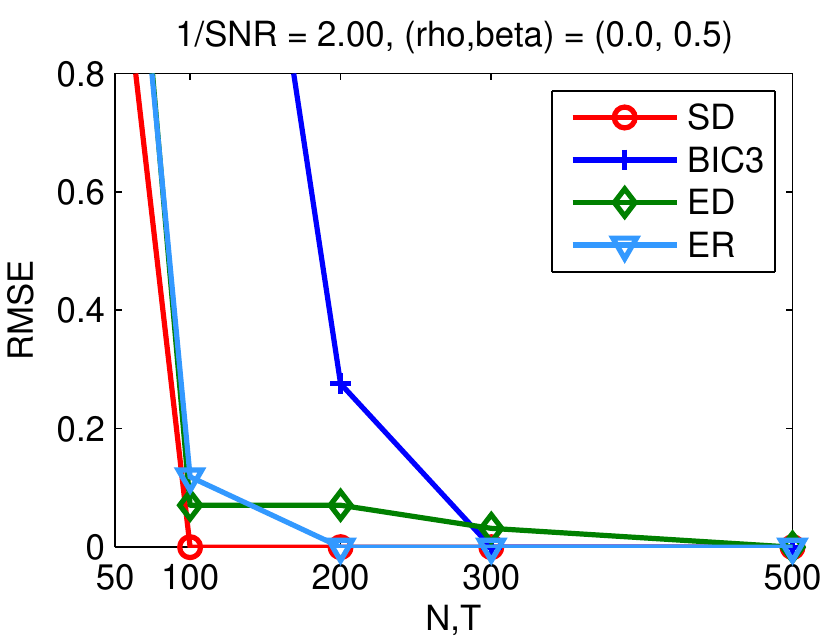}&
\includegraphics[width=0.8\linewidth]{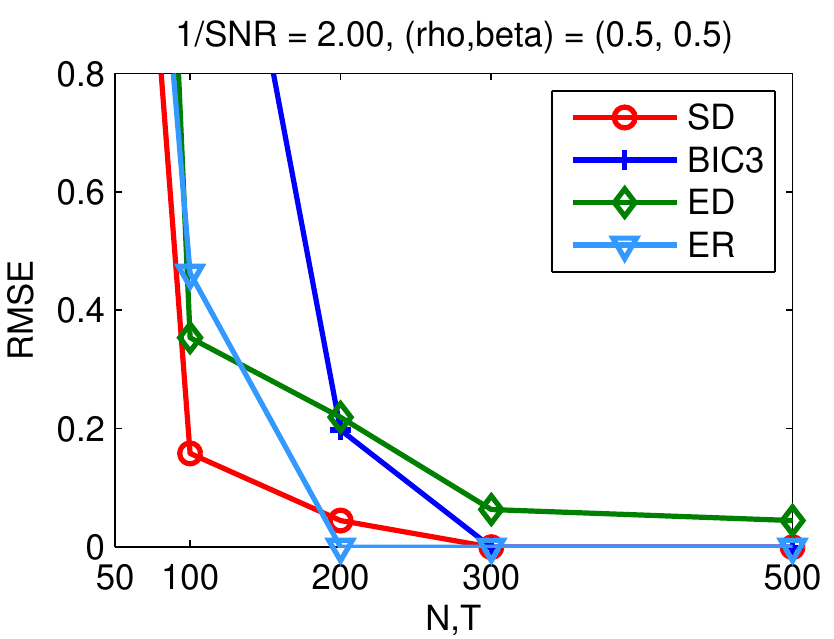}\\
\hline
\end{tabular}}
\caption[Root mean square errors (RMSE) for the estimated number of factors with respect to $N$]{Root mean square errors (RMSE) for the estimated number of factors with respect to $N$. Each plot is generated with different noise level: 1/SNR=0.1, 0,25, 0.5, 0.75, 1, and 2. Here we set $p=4$, $T=N$. The residuals have correlation structure: cross-correlations $(\rho,\beta)=(0,0.5)$ (left) and auto- and cross-correlations $(\rho, \beta)=(0.5,0.5)$ (right). We set $J = N/10$. Our estimator converges sufficiently well for $N \geq 100$, regardless of signal-to-noise ratios.}
\label{fig::MC_NT}
\end{figure}

Figure \ref{fig::MC_NT} reports the convergence speed of estimators with respect to sample sizes. The true number of factors is 4, and we set $T=N$. Residuals have correlations, as $(\rho,\beta)=(0, 0.5)$ or $(0.5,0.5)$, and $J=N/10$. As seen from the figure, it is clear that the estimators are generally converging to the true number of factors as $N$ and $T$ become large. When the amount of noise is small, BIC3 and ER converges the fastest. However, as the noise level increases, our estimator outperforms others especially with small sample sizes.

\begin{figure}[!htbp]
\centering
\scalebox{0.80}{
\begin{tabular}{| >{\centering\arraybackslash}m{0.9in} | >{\centering\arraybackslash}m{2.4in}  >{\centering\arraybackslash}m{2.4in}  |}
\hline
& {N = 100} & {N = 200} \\
\hline
\texttt{no correlation} (0,0) &
\includegraphics[width=1\linewidth]{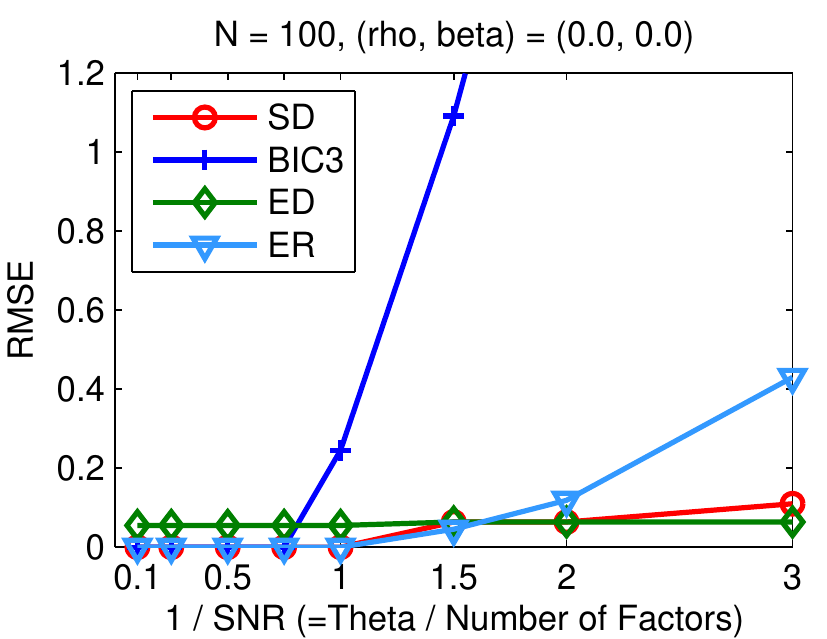}&
\includegraphics[width=1\linewidth]{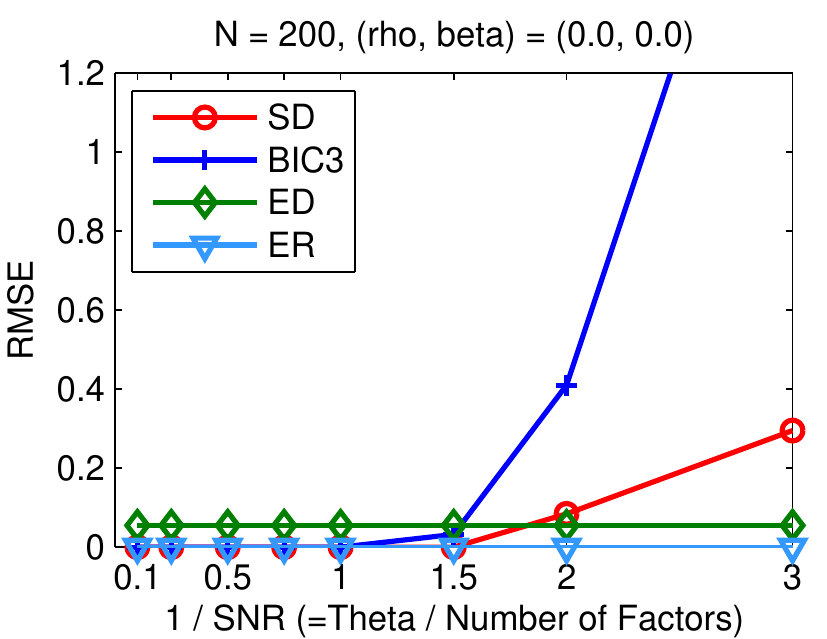}\\
\texttt{auto correlation} (0.5, 0) &
\includegraphics[width=1\linewidth]{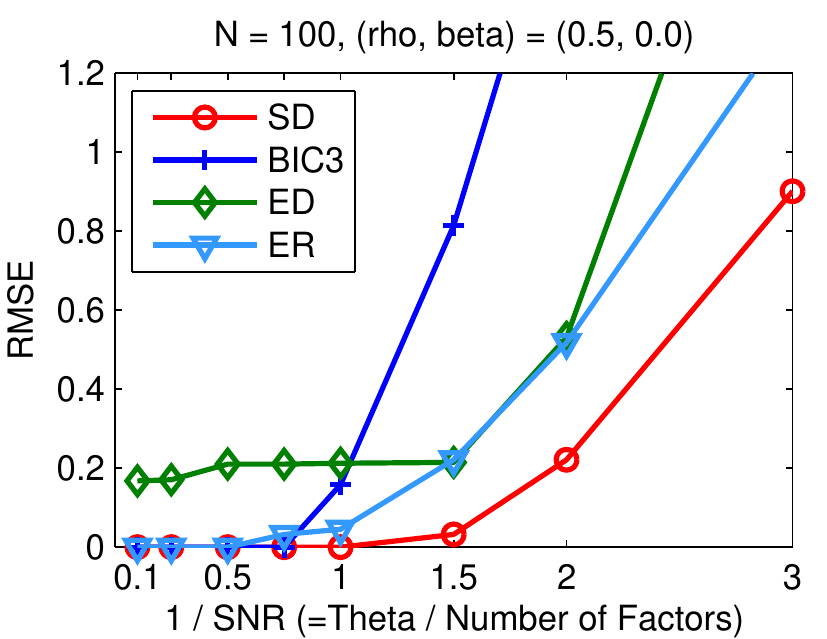}&
\includegraphics[width=1\linewidth]{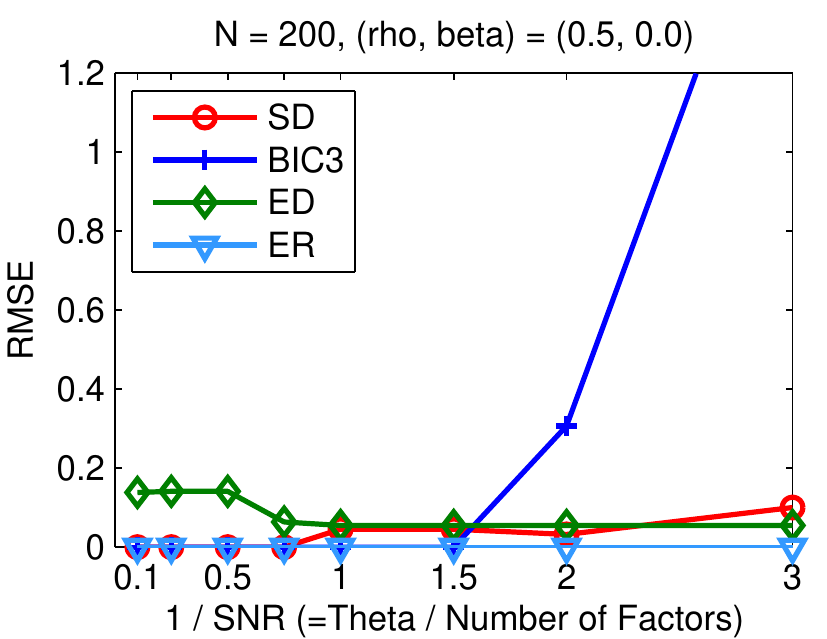}\\
\texttt{cross correlation} (0, 0.5) &
\includegraphics[width=1\linewidth]{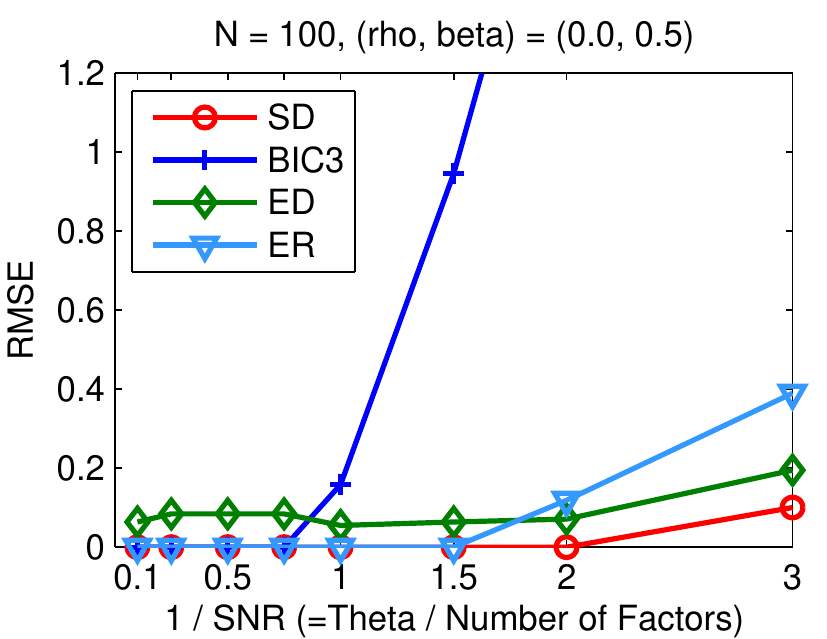}&
\includegraphics[width=1\linewidth]{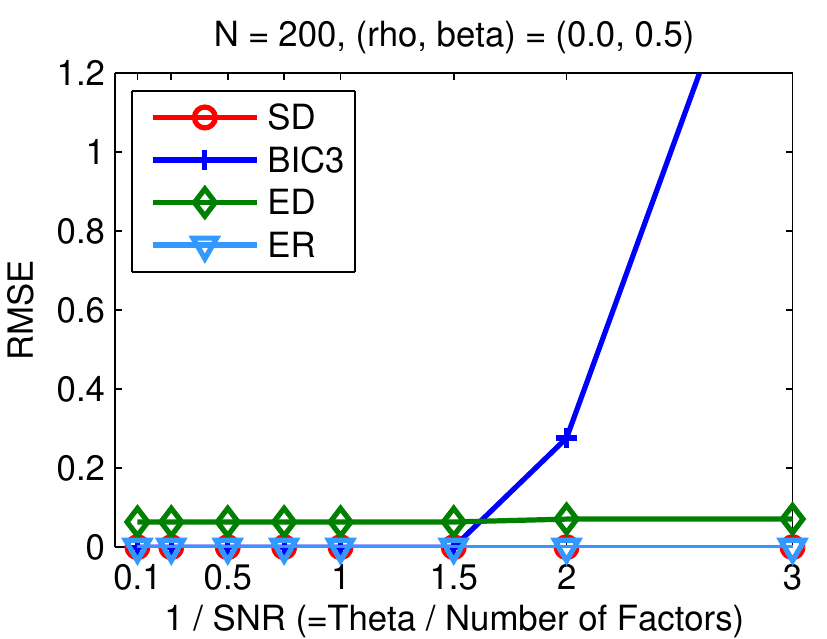}\\
\texttt{auto- and cross- correlation (0.5, 0.5)} &
\includegraphics[width=1\linewidth]{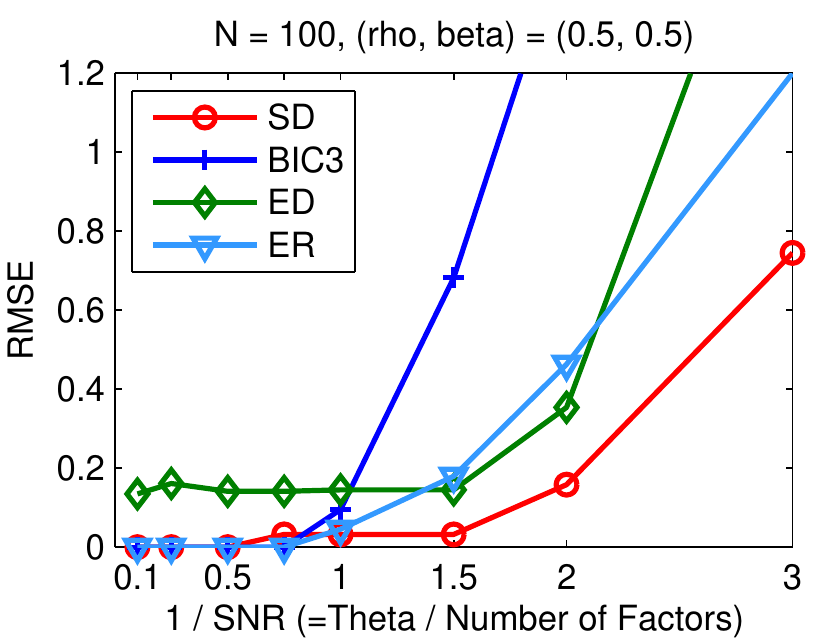}&
\includegraphics[width=1\linewidth]{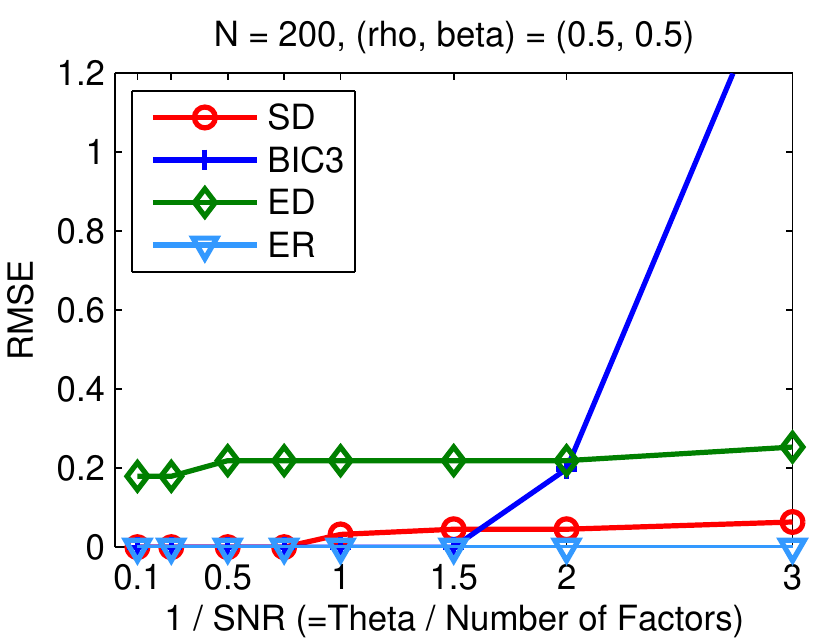}\\
\hline
\end{tabular}}
\caption[Root mean square errors (RMSE) with respect to noise level (1/SNR)]{Root mean square errors (RMSE) with respect to noise level (1/SNR).  Each column shows results from different $N=T$ = 100 (left) and 200 (right). Each row represents different correlation structures in residuals: From top to bottom, $(\rho, \beta )=(0,0), (0.5,0), (0,0.5), (0.5,0.5)$. The estimation error is increasing when the noise amount of residual becomes larger. Note that for comparatively small sample size $(N\leq 100)$, our estimation method is more robust to residual noises than other methods.}
\label{fig::MC_SNR}
\end{figure}

This result is also reflected in Figure \ref{fig::MC_SNR}, where the graphs of RMSE are drawn with respect to the noise level. Clearly, higher noise levels inhibit the estimation precisions. In addition, it is easy to observe that SD is less sensitive to noise amount than other methods, especially for smaller sample size $(N=100)$. For larger sample size ($N=200$), ER shows the best performance, followed by SD which is still stable from noise disturbance. We also discovered that the considered cross-correlation structure is less affected than auto-correlation structure from increasing noise amounts. In the meantime, BIC3 is the most vulnerable to noise levels.

\subsubsection{Presence of weak factors}
\label{sec::weak}
Detecting weak factors is generally harder than detecting strong factors. In this section, similar to the experiments in \cite{AH2013}, we study the influence of weak factors on the estimated number of factors.

\begin{figure}[!htbp]
\centering
\scalebox{0.85}{
\begin{tabular}{| >{\centering\arraybackslash}m{0.9in} | >{\centering\arraybackslash}m{2.4in}  >{\centering\arraybackslash}m{2.4in}  |}
\hline
& 3 (out of 4) Weak factors & 4 (out of 4) Weak factors \\
\hline
\texttt{N=50} &
\includegraphics[width=1\linewidth]{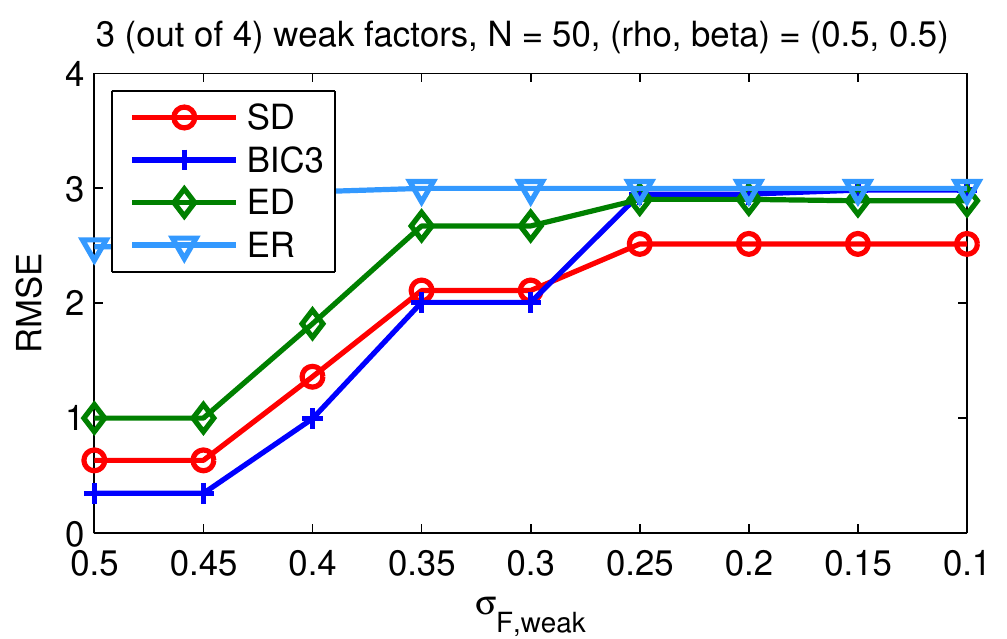}&
\includegraphics[width=1\linewidth]{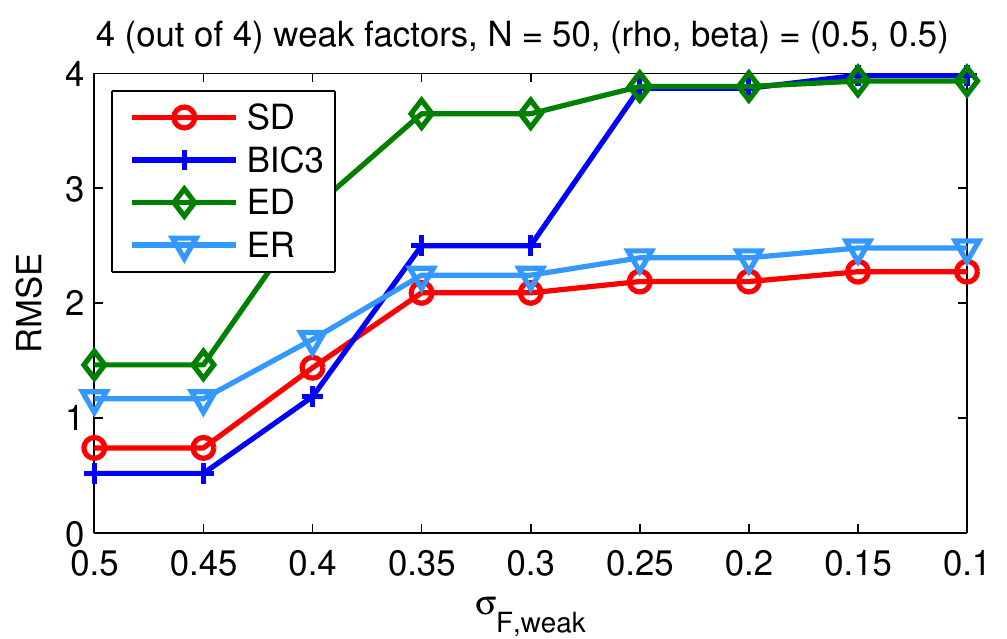}\\
\texttt{N=100}&
\includegraphics[width=1\linewidth]{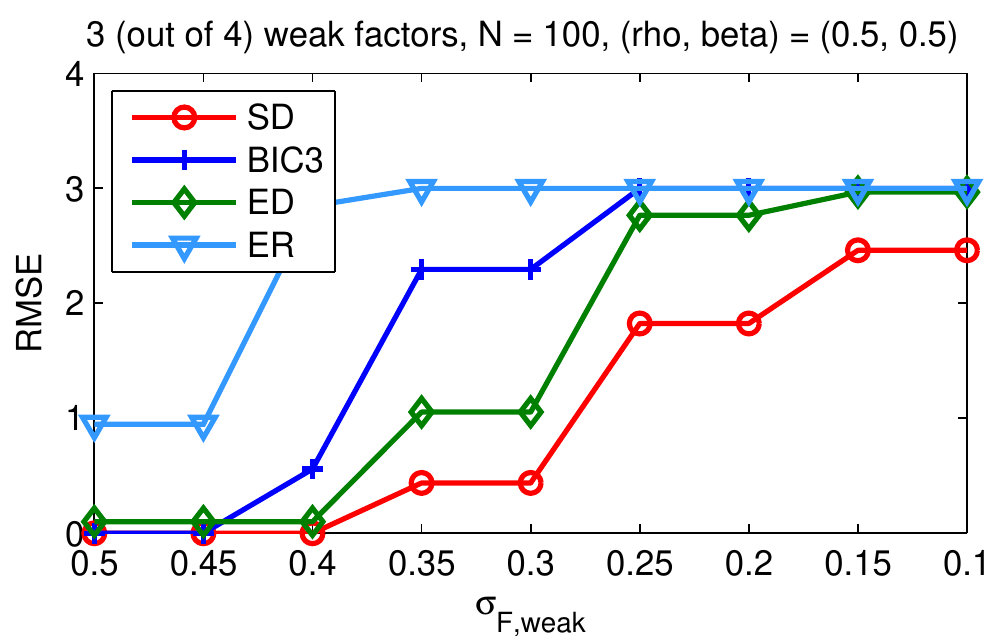}&
\includegraphics[width=1\linewidth]{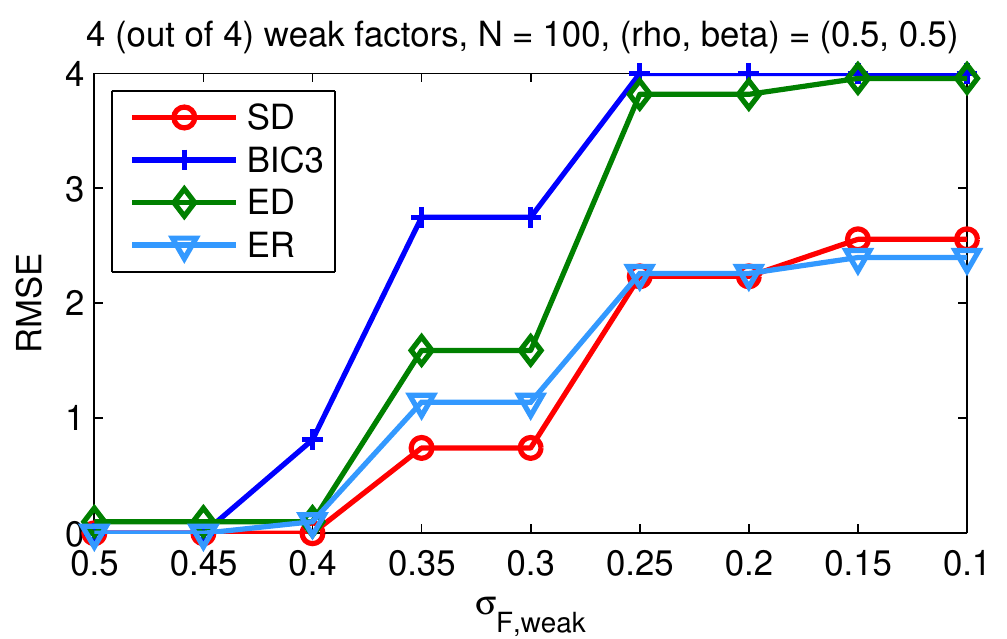}\\
\texttt{N=200}&
\includegraphics[width=1\linewidth]{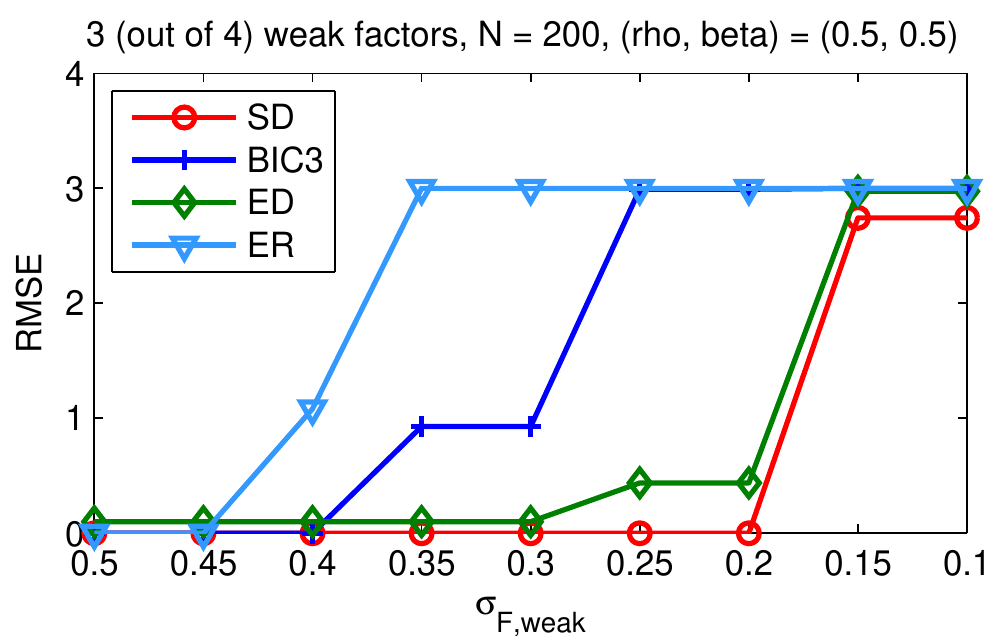}&
\includegraphics[width=1\linewidth]{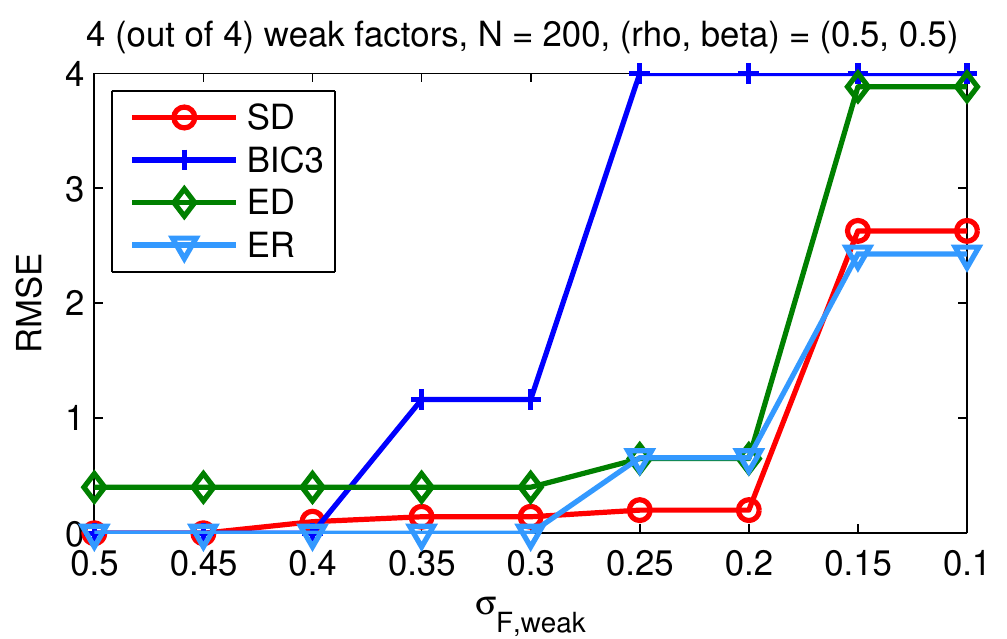}\\
\texttt{N=500} &
\includegraphics[width=1\linewidth]{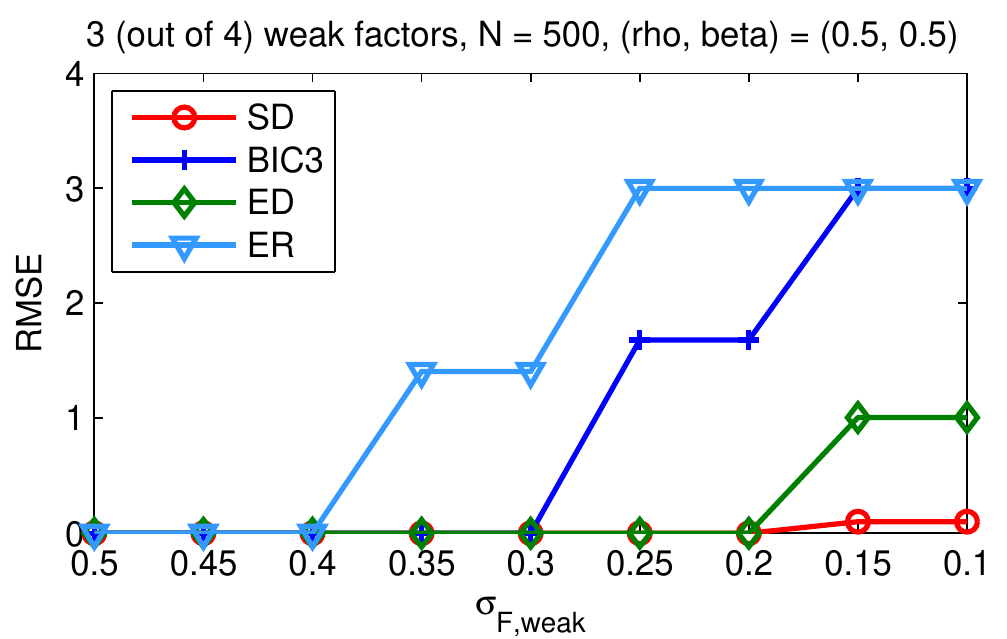}&
\includegraphics[width=1\linewidth]{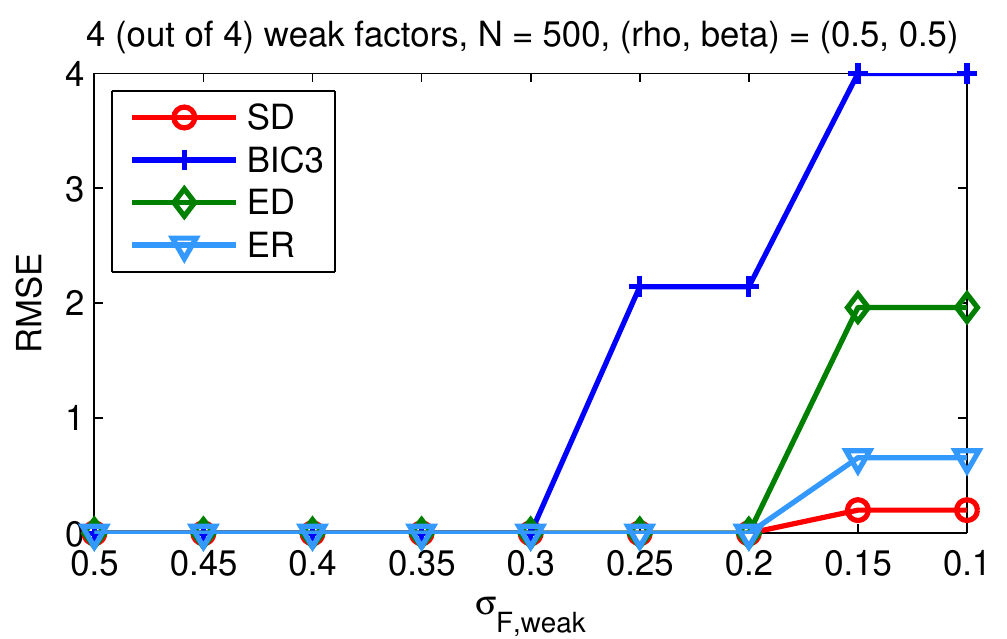}\\
\hline
\end{tabular}}
\caption[Root mean square errors (RMSE) with respect to variance of weak factors]{Root mean square errors (RMSE) with respect to variance of weak factors. The true number of factors is 4. The left column represents when 3 factors are weak, and the right column is for when all 4 factors are weak. Sample sizes are also varied: $N=50, 100, 200, 500$, and $T=N$. Residuals have correlation structures: $(\rho, \beta )=(0.5,0.5)$, and $J=N/10$. The ability of SD to detect the weak factors are significantly better than other methods.}
\label{fig::MC_weak}
\end{figure}

To construct weak factors, we reduce the variance of $f_{jt}$ in Eq.\ref{eq::synthetic} to be less than one: $f^{weak}_{jt}\sim N(0, \sigma_{weak}^2)$, with $\sigma_{weak} < 1$. We set four true factors, and consider two cases: (1) the case where all four factors are weak and (2) the case where only three factors are weak. The performance with weak factors is compared in Figure \ref{fig::MC_weak}. Clearly, if the factors get weaker (smaller $\sigma_{weak}$), it becomes harder to detect the those weak factors, which results in increasing estimation errors as presented in the figure. In addition, if there is one stronger factor and several weak factors, it is generally more difficult to distinguish weak ones. This explains the fact that the overall RMSE values on the left column is larger than that on the right column.

More importantly, this figure provides evidence that our method (SD) has more powerful ability to identify weak factors, compared to other methods, from all of the considered cases. There are several possible explanations for this result. Note that the spectral distance measure we consider has larger weights for the spikes in the spectrum. Therefore, if the eigenvalues corresponding to weak factors are not diverging much and staying outside the bulk, our algorithm is likely to detect them as factors. Besides, the control parameter $b$ allows to amplify the resolution of detecting those weaker factors. On the other hand, other methods do not take into account this mechanism in their algorithms.

\section{Applications to real data}
\label{sec::realdata}
In this section, we apply the proposed methods to market data. Daily returns of 378 stocks\footnote{We consider stocks who have survived persistently in the entire period.} in S\&P500 between 2000-2015 are used.

Instead of taking the entire time range at once, we use a certain length of estimation window, and move the window one day at a time. There is an overlap in the data contained in consecutive windows, which enables us to track the temporal evolution of number of factors and correlation structure of residuals. The estimation with moving windows produces $\hat p$ and $\hat b$ for each day, giving the time-series of estimators.

\subsection{Static experiment}
\begin{figure}[!htbp]
\centering
\begin{tabular}{cc}
\includegraphics[width=0.4\linewidth]{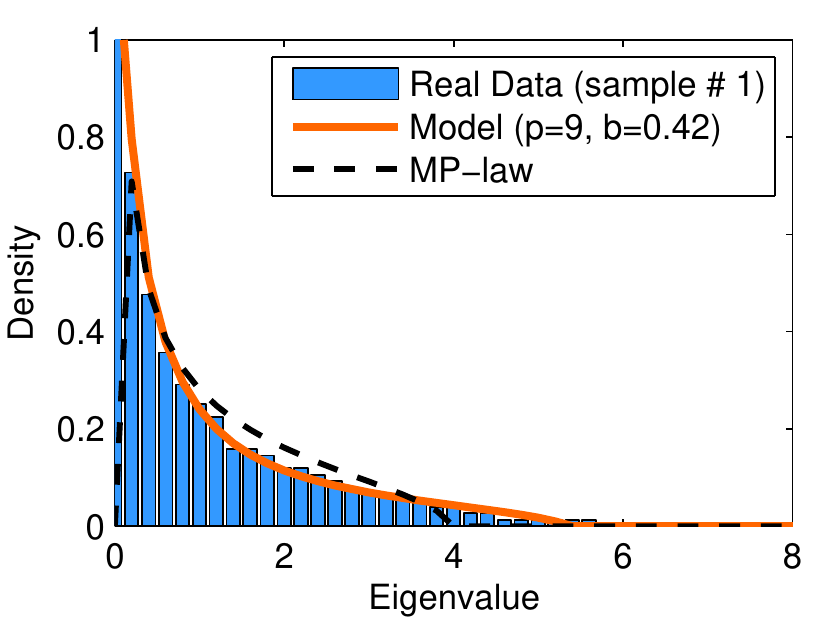}&
\includegraphics[width=0.4\linewidth]{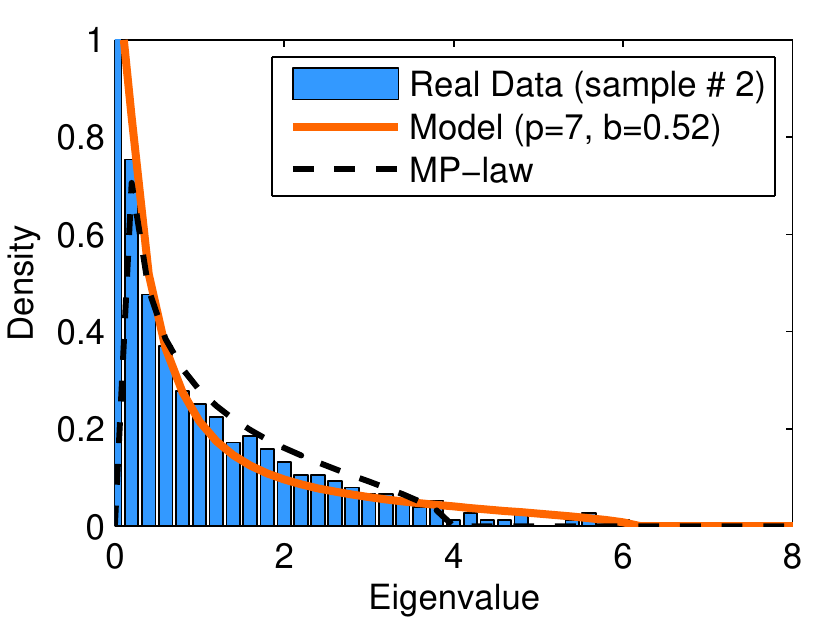}\\
\includegraphics[width=0.4\linewidth]{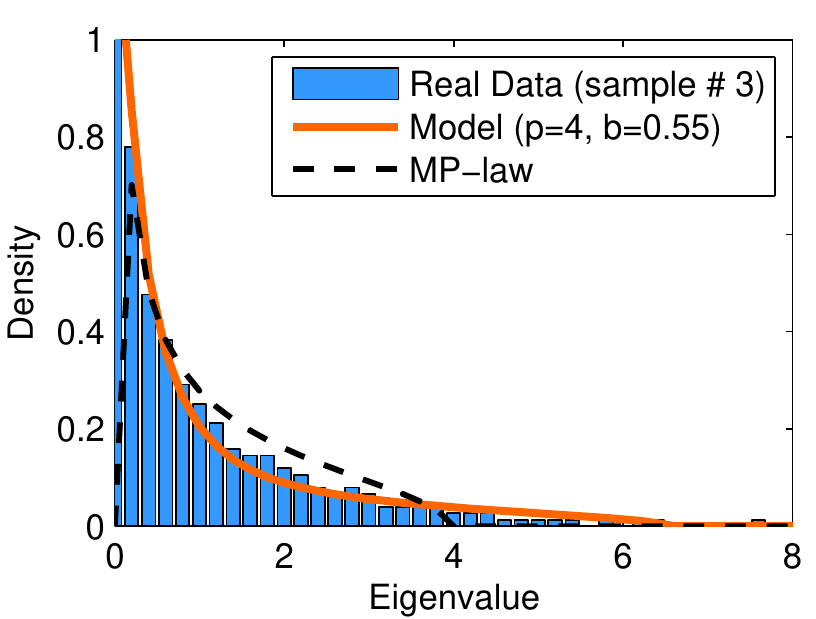}&
\includegraphics[width=0.4\linewidth]{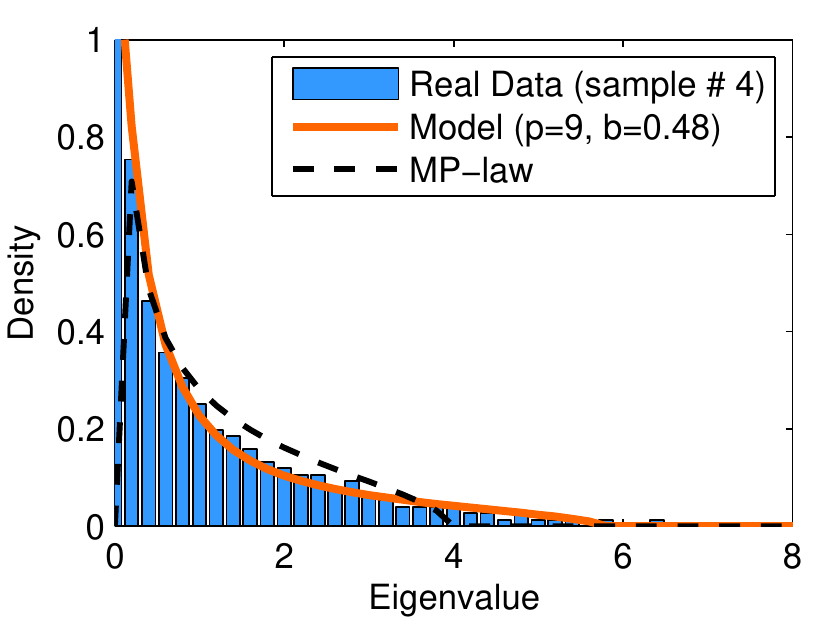}\\
\end{tabular}
\caption[Fit of simplified model to real data]{Fit of simplified model to real data. Four datasets are randomly selected from the year of 2001, 2005, 2008, and 2011. The model density with estimated $p$ and $b$ generally fits the spectrum of residuals well. For comparison, Marchenko-Pastur (MP) law for the correlation matrix is plotted.}
\label{fig::fitRealModel}
\end{figure}

Before discussing dynamics of estimated parameters, we first check how well the simplified model can fit the residuals from real data. In Figure \ref{fig::fitRealModel}, we show several sample fitted results. Four random days are selected in the year of 2001, 2005, 2008, and 2011 and the factor model estimation using the simplified model is applied to each data. Note that the estimated $p$ and $b$'s are different for different data, but each density from estimated model explains well the eigenvalue distribution of correlation matrix of real residuals, compared to corresponding Marchenko-Pastur law.

\subsection{Dynamic experiment: implications of $\hat p$}
\begin{figure}[!htbp]
\centering
\begin{tabular}{c}
\includegraphics[width=1\linewidth]{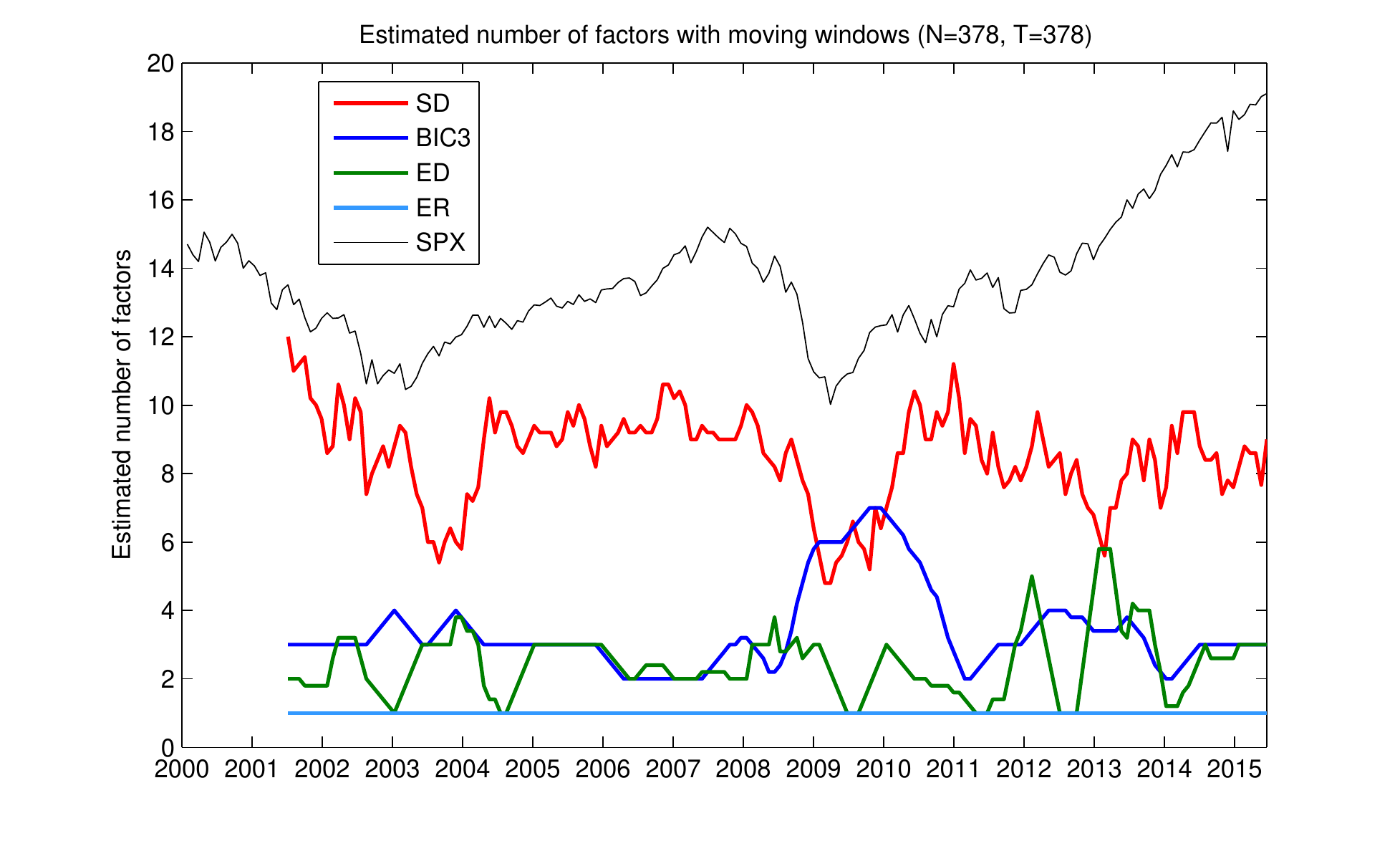}
\end{tabular}
\caption[Estimated number of factors are compared with other methods]{Estimated number of factors are compared with other methods. For real data, the results vary depending on methods. Overall, our estimator (SD) is larger than others. Eigenvalue Ratio (ER) method by \cite{AH2013} gives always one factor throughout the whole investigated period. Information criteria based method (BIC3) by \cite{BaiNg2002} and eigenvalue difference method (ED) by \cite{Onatski2010} display almost opposite results during crisis (2008-2009) The estimated number of factors from BIC3 is increasing, while that of ED decreases at the same time. The two provide the same number of factors during 2005-2006. The investigated stocks are $N=378$, time window is $T=378$ business days. Note that the estimated values are available only from mid-2001, because of the length of moving window for estimations.}
\label{fig::timeseries_2}
\end{figure}

Repetitive estimation procedures with moving windows generate time-series of $\hat p$. In order to evaluate the performance of estimated number of factors $\hat p$, we compare those from other methods. Figure \ref{fig::timeseries_2} reports the estimators.
It is clear that our estimator is between 4 and 12, changing in time, which is mostly larger than others that display 1 to 7 factors. The most likely explanation is that our method identifies several weak factors in addition to strong factors\footnote{Note that the Fama-French three factor model \cite{FF1992} has been used widely in explaining the returns of equity securities. But as reported \cite{FKL2016}, testing Fama-French model is more often to be rejected when using the daily data, compared to  the monthly data, due to a larger volatility of the unexplained factor components.}, as we already have seen from the Monte carlo results in Section \ref{sec::weak}.

We investigate what those factors actually consist of. To do this, we examine the components in the eigenvectors corresponding to top eigenvalues of correlation matrix of returns. As seen from Table \ref{table::eigenvector}, the components in the first eigenvector are very uniformly contributed, which indicates the market mode. From the second, each eigenvector corresponds to business sector. For example, during 2004-2005, the major three factors are Energy, Financial REITs, and Information Technology. However, during 2008-2009, Energy sector takes the second and the third factors, and Financials are the fourth factor. In any cases, although the principal component factors are constructed from purely statistical procedure, they are closely related to business sectors.

Meanwhile, it is interesting to note that eigenvalue ratio (ER) methods gives one factor all the time. Information criteria-based method (BIC3) shows the nearly opposite behaviors to other methods, especially estimating more factors in crisis. We have not found a clear reason for that.
% \usepackage{multirow}
\begin{table}[]
\centering
\scalebox{0.30}{
\begin{tabular}{|c|c|c|c|c|c|c|c|c|c|c|c|}
\hline
\multicolumn{6}{|c|}{2004-2005}                                                                                                                                                                                                      & \multicolumn{6}{c|}{2008-2009}                                                                                                                                                                            \\ \hline
\textbf{\#}          & \textbf{Eigenvalue}      & \textbf{Company}                                                 & \textbf{Sector}                                                  & \textbf{Components} & \textbf{Contributions} & \textbf{\#}          & \textbf{Eigenvalue}      & \textbf{Company}                                                                & \textbf{Sector}        & \textbf{Components} & \textbf{Contributions} \\ \hline
\multirow{10}{*}{1}  & \multirow{10}{*}{0.2485} & \begin{tabular}[c]{@{}c@{}}PPG Industries\end{tabular}       & Chemical                                                         & 0.076               & 0.40\%                 & \multirow{10}{*}{1}  & \multirow{10}{*}{0.5096} & \begin{tabular}[c]{@{}c@{}}E.I. Du Pont De Nemours \& Co.\end{tabular} & Chemical               & 0.064               & 0.33\%                 \\ \cline{3-6} \cline{9-12}
                     &                          & T. Rowe Price Group                                              & Financials                                                       & 0.074               & 0.39\%                 &                      &                          & Walt Disney Company                                                             & Media                  & 0.062               & 0.32\%                 \\ \cline{3-6} \cline{9-12}
                     &                          & Northern Trust Corp.                                             & Financials                                                       & 0.073               & 0.39\%                 &                      &                          & Illinois Tool Works Inc                                                         & Machinery              & 0.062               & 0.32\%                 \\ \cline{3-6} \cline{9-12}
                     &                          & Emerson Electric Company                                         & Industrials                                                      & 0.073               & 0.39\%                 &                      &                          & PPG Industries                                                                  & Chemical               & 0.062               & 0.32\%                 \\ \cline{3-6} \cline{9-12}
                     &                          & Praxair Inc.                                                     & Materials                                                        & 0.073               & 0.38\%                 &                      &                          & Franklin Resources                                                              & Financial              & 0.062               & 0.32\%                 \\ \cline{3-6} \cline{9-12}
                     &                          & SunTrust Banks                                                   & Financials                                                       & 0.072               & 0.38\%                 &                      &                          & United Technologies                                                             & Industrials            & 0.062               & 0.32\%                 \\ \cline{3-6} \cline{9-12}
                     &                          & PACCAR Inc.                                                      & Industrials                                                      & 0.072               & 0.38\%                 &                      &                          & Stanley Black \& Decker                                                         & Consumer Discretionary & 0.062               & 0.32\%                 \\ \cline{3-6} \cline{9-12}
                     &                          & BB\&T Corporation                                                & Financials                                                       & 0.072               & 0.38\%                 &                      &                          & T. Rowe Price Group                                                             & Financials             & 0.061               & 0.32\%                 \\ \cline{3-6} \cline{9-12}
                     &                          & M\&T Bank Corp.                                                  & Financials                                                       & 0.072               & 0.38\%                 &                      &                          & Equifax Inc.                                                                    & Financials             & 0.061               & 0.32\%                 \\ \cline{3-6} \cline{9-12}
                     &                          & Realty Income Corporation                                        & Real Estate                                                      & 0.071               & 0.38\%                 &                      &                          & Invesco Ltd.                                                                    & Financials             & 0.061               & 0.31\%                 \\ \hline
\multirow{10}{*}{2}  & \multirow{10}{*}{0.0571} & Apache Corporation                                               & Energy                                                           & 0.167               & 1.26\%                 & \multirow{10}{*}{2}  & \multirow{10}{*}{0.0499} & Noble Energy Inc                                                                & Energy                 & 0.112               & 0.72\%                 \\ \cline{3-6} \cline{9-12}
                     &                          & Chesapeake Energy                                                & Energy                                                           & 0.165               & 1.25\%                 &                      &                          & Apache Corp                                                                     & Energy                 & 0.112               & 0.72\%                 \\ \cline{3-6} \cline{9-12}
                     &                          & Newfield Exploration Co                                          & Energy                                                           & 0.163               & 1.23\%                 &                      &                          & Southwestern Energy Company                                                     & Energy                 & 0.109               & 0.70\%                 \\ \cline{3-6} \cline{9-12}
                     &                          & Diamond Offshore Drilling                                        & Energy                                                           & 0.161               & 1.22\%                 &                      &                          & Anadarko Petroleum Corp                                                         & Energy                 & 0.108               & 0.69\%                 \\ \cline{3-6} \cline{9-12}
                     &                          & EOG Resources                                                    & Energy                                                           & 0.161               & 1.22\%                 &                      &                          & Entergy Corp                                                                    & Energy                 & 0.108               & 0.69\%                 \\ \cline{3-6} \cline{9-12}
                     &                          & Noble Energy Inc                                                 & Energy                                                           & 0.161               & 1.22\%                 &                      &                          & National Oilwell Varco Inc.                                                     & Energy                 & 0.106               & 0.68\%                 \\ \cline{3-6} \cline{9-12}
                     &                          & Ensco plc                                                        & Energy                                                           & 0.161               & 1.22\%                 &                      &                          & Diamond Offshore Drilling                                                       & Energy                 & 0.105               & 0.68\%                 \\ \cline{3-6} \cline{9-12}
                     &                          & Hess Corporation                                                 & Energy                                                           & 0.160               & 1.21\%                 &                      &                          & Cabot Oil \& Gas                                                                & Energy                 & 0.105               & 0.67\%                 \\ \cline{3-6} \cline{9-12}
                     &                          & \begin{tabular}[c]{@{}c@{}}Occidental Petroleum\end{tabular} & Energy                                                           & 0.158               & 1.20\%                 &                      &                          & \begin{tabular}[c]{@{}c@{}}Chesapeake Energy\end{tabular}                   & Energy                 & 0.104               & 0.67\%                 \\ \cline{3-6} \cline{9-12}
                     &                          & Transocean                                                       & Energy                                                           & 0.158               & 1.19\%                 &                      &                          & Kimco Realty                                                                    & Financials             & (0.104)             & (0.67\%)               \\ \hline
\multirow{10}{*}{3}  & \multirow{10}{*}{0.0345} & Boston Properties                                                & Financials REITs                                                 & (0.161)             & (1.12\%)               & \multirow{10}{*}{3}  & \multirow{10}{*}{0.0304} & Freeport-Mcmoran Inc                                                            & Mining                 & 0.139               & 0.90\%                 \\ \cline{3-6} \cline{9-12}
                     &                          & Macerich                                                         & Financials REITs                                                 & (0.159)             & (1.11\%)               &                      &                          & Consol Energy                                                                   & Energy                 & 0.133               & 0.86\%                 \\ \cline{3-6} \cline{9-12}
                     &                          & Simon Property Group Inc                                         & Financials REITs                                                 & (0.156)             & (1.09\%)               &                      &                          & Southern Co.                                                                    & Energy                 & (0.128)             & (0.83\%)               \\ \cline{3-6} \cline{9-12}
                     &                          & Public Storage                                                   & Financials REITs                                                 & (0.151)             & (1.05\%)               &                      &                          & Cameron International                                                           & Energy                 & 0.126               & 0.82\%                 \\ \cline{3-6} \cline{9-12}
                     &                          & Kimco Realty                                                     & Financials REITs                                                 & (0.148)             & (1.03\%)               &                      &                          & Newfield Exploration Co                                                         & Energy                 & 0.124               & 0.80\%                 \\ \cline{3-6} \cline{9-12}
                     &                          & Vornado Realty Trust                                             & Financials REITs                                                 & (0.147)             & (1.02\%)               &                      &                          & Consolidated Edison                                                             & Utilities              & (0.118)             & (0.77\%)               \\ \cline{3-6} \cline{9-12}
                     &                          & SL Green Realty                                                  & Financials REITs                                                 & (0.147)             & (1.02\%)               &                      &                          & Duke Energy                                                                     & Energy                 & (0.117)             & (0.76\%)               \\ \cline{3-6} \cline{9-12}
                     &                          & AvalonBay Communities, Inc.                                      & Financials REITs                                                 & (0.146)             & (1.02\%)               &                      &                          & Campbell Soup                                                                   & Consumer Staples       & (0.116)             & (0.76\%)               \\ \cline{3-6} \cline{9-12}
                     &                          & Equity Residential                                               & Financials REITs                                                 & (0.146)             & (1.02\%)               &                      &                          & Range Resources Corp.                                                           & Energy                 & 0.116               & 0.75\%                 \\ \cline{3-6} \cline{9-12}
                     &                          & HCP Inc.                                                         & Financials REITs                                                 & (0.144)             & (1.00\%)               &                      &                          & Abbott Laboratories                                                             & Health Care            & (0.113)             & (0.73\%)               \\ \hline
\multirow{10}{*}{4}  & \multirow{10}{*}{0.0192} & Altera Corp                                                      & Information Technology                                           & 0.163               & 1.09\%                 & \multirow{10}{*}{4}  & \multirow{10}{*}{0.0189} & KeyCorp                                                                         & Financials             & (0.195)             & (1.33\%)               \\ \cline{3-6} \cline{9-12}
                     &                          & Xilinx Inc                                                       & Information Technology                                           & 0.154               & 1.03\%                 &                      &                          & Morgan Stanley                                                                  & Financials             & (0.187)             & (1.27\%)               \\ \cline{3-6} \cline{9-12}
                     &                          & Lam Research                                                     & Information Technology                                           & 0.146               & 0.98\%                 &                      &                          & \begin{tabular}[c]{@{}c@{}}XL Capital\end{tabular}                          & Financials             & (0.173)             & (1.18\%)               \\ \cline{3-6} \cline{9-12}
                     &                          & Broadcom Corporation                                             & Information Technology                                           & 0.145               & 0.97\%                 &                      &                          & Hartford Financial Svc.Gp.                                                      & Financials             & (0.173)             & (1.18\%)               \\ \cline{3-6} \cline{9-12}
                     &                          & KLA-Tencor Corp.                                                 & Information Technology                                           & 0.144               & 0.96\%                 &                      &                          & SunTrust Banks                                                                  & Financials             & (0.170)             & (1.16\%)               \\ \cline{3-6} \cline{9-12}
                     &                          & Microchip Technology                                             & Information Technology                                           & 0.143               & 0.96\%                 &                      &                          & Citigroup Inc.                                                                  & Financials             & (0.159)             & (1.08\%)               \\ \cline{3-6} \cline{9-12}
                     &                          & Skyworks Solutions                                               & Information Technology                                           & 0.143               & 0.96\%                 &                      &                          & Bank of America Corp                                                            & Financials             & (0.156)             & (1.06\%)               \\ \cline{3-6} \cline{9-12}
                     &                          & Applied Materials Inc                                            & Information Technology                                           & 0.122               & 0.82\%                 &                      &                          & Zions Bancorp                                                                   & Financials             & (0.150)             & (1.02\%)               \\ \cline{3-6} \cline{9-12}
                     &                          & Linear Technology Corp.                                          & Information Technology                                           & 0.120               & 0.80\%                 &                      &                          & Goldman Sachs Group                                                             & Financials             & (0.148)             & (1.01\%)               \\ \cline{3-6} \cline{9-12}
                     &                          & Motorola Solutions Inc.                                          & Information Technology                                           & 0.118               & 0.79\%                 &                      &                          & Comerica Inc.                                                                   & Financials             & (0.138)             & (0.94\%)               \\ \hline
\multirow{10}{*}{5}  & \multirow{10}{*}{0.0164} & TJX Companies Inc.                                               & Consumer Discretionary                                           & (0.159)             & (1.06\%)               & \multirow{10}{*}{5}  & \multirow{10}{*}{0.0137} & Realty Income Corporation                                                       & Financials REITs       & 0.157               & 1.02\%                 \\ \cline{3-6} \cline{9-12}
                     &                          & Urban Outfitters                                                 & Consumer Discretionary                                           & (0.130)             & (0.87\%)               &                      &                          & Equity Residential                                                              & Financials REITs       & 0.147               & 0.96\%                 \\ \cline{3-6} \cline{9-12}
                     &                          & Nordstrom                                                        & Consumer Discretionary                                           & (0.126)             & (0.84\%)               &                      &                          & Public Storage                                                                  & Financials REITs       & 0.147               & 0.96\%                 \\ \cline{3-6} \cline{9-12}
                     &                          & Applied Materials Inc                                            & Information Technology                                           & 0.123               & 0.81\%                 &                      &                          & Coca-Cola Enterprises'                                                          & Consumer Staples       & (0.144)             & (0.94\%)               \\ \cline{3-6} \cline{9-12}
                     &                          & Prologis                                                         & Financials REITs                                                 & (0.116)             & (0.77\%)               &                      &                          & Welltower Inc.                                                                  & Financials REITs       & 0.142               & 0.93\%                 \\ \cline{3-6} \cline{9-12}
                     &                          & Lowe's Cos.                                                      & Consumer Discretionary                                           & (0.116)             & (0.77\%)               &                      &                          & Priceline.com Inc                                                               & Consumer Discretionary & 0.133               & 0.87\%                 \\ \cline{3-6} \cline{9-12}
                     &                          & Broadcom                                                         & Information Technology                                           & 0.115               & 0.76\%                 &                      &                          & Vornado Realty Trust                                                            & Financials             & 0.130               & 0.85\%                 \\ \cline{3-6} \cline{9-12}
                     &                          & Target Corp.                                                     & Consumer Discretionary                                           & (0.110)             & (0.73\%)               &                      &                          & Universal Health Services                                                       & Health Care            & (0.121)             & (0.79\%)               \\ \cline{3-6} \cline{9-12}
                     &                          & Ross Stores                                                      & Consumer Discretionary                                           & (0.110)             & (0.73\%)               &                      &                          & Simon Property Group Inc                                                        & Financials             & 0.119               & 0.78\%                 \\ \cline{3-6} \cline{9-12}
                     &                          & Amazon.com Inc                                                   & Consumer Discretionary                                           & (0.109)             & (0.72\%)               &                      &                          & AvalonBay Communities, Inc.                                                     & Financials             & 0.117               & 0.77\%                 \\ \hline
\multirow{10}{*}{6}  & \multirow{10}{*}{0.013}  & Parker-Hannifin                                                  & Industrials                                                      & 0.136               & 0.86\%                 & \multirow{10}{*}{6}  & \multirow{10}{*}{0.0109} & \begin{tabular}[c]{@{}c@{}}Broadcom Corporation\end{tabular}                & Information Technology & (0.213)             & (1.39\%)               \\ \cline{3-6} \cline{9-12}
                     &                          & Dow Chemical                                                     & Materials                                                        & 0.135               & 0.85\%                 &                      &                          & Altera Corp                                                                     & Information Technology & (0.164)             & (1.07\%)               \\ \cline{3-6} \cline{9-12}
                     &                          & Cummins Inc.                                                     & Industrials                                                      & 0.130               & 0.82\%                 &                      &                          & Apple Inc.                                                                      & Information Technology & (0.147)             & (0.96\%)               \\ \cline{3-6} \cline{9-12}
                     &                          & Eastman Chemical                                                 & Materials                                                        & 0.122               & 0.77\%                 &                      &                          & Microchip Technology                                                            & Information Technology & (0.144)             & (0.94\%)               \\ \cline{3-6} \cline{9-12}
                     &                          & FMC Corporation                                                  & Materials                                                        & 0.121               & 0.76\%                 &                      &                          & NetApp                                                                          & Information Technology & (0.142)             & (0.93\%)               \\ \cline{3-6} \cline{9-12}
                     &                          & American International Group, Inc.                               & Financials                                                       & (0.120)             & (0.76\%)               &                      &                          & Xilinx Inc                                                                      & Information Technology & (0.140)             & (0.92\%)               \\ \cline{3-6} \cline{9-12}
                     &                          & \begin{tabular}[c]{@{}c@{}}Stryker Corp.\end{tabular}        & Health Care                                                      & (0.119)             & (0.75\%)               &                      &                          & \begin{tabular}[c]{@{}c@{}}F5 Networks Inc.\end{tabular}                    & Information Technology & (0.139)             & (0.91\%)               \\ \cline{3-6} \cline{9-12}
                     &                          & Ingersoll-Rand PLC                                               & Industrials                                                      & 0.113               & 0.71\%                 &                      &                          & Intel Corp.                                                                     & Information Technology & (0.139)             & (0.91\%)               \\ \cline{3-6} \cline{9-12}
                     &                          & Deere \& Co.                                                     & Industrials                                                      & 0.112               & 0.70\%                 &                      &                          & Citrix Systems                                                                  & Information Technology & (0.136)             & (0.89\%)               \\ \cline{3-6} \cline{9-12}
                     &                          & \begin{tabular}[c]{@{}c@{}}PPG Industries\end{tabular}       & Materials                                                        & 0.110               & 0.69\%                 &                      &                          & Nvidia Corporation                                                              & Information Technology & (0.125)             & (0.82\%)               \\ \hline
\multirow{10}{*}{7}  & \multirow{10}{*}{0.012}  & Nordstrom                                                        & Consumer Discretionary                                           & (0.152)             & (0.98\%)               & \multirow{10}{*}{7}  & \multirow{10}{*}{0.0097} & DaVita Inc.                                                                     & Health Care            & 0.146               & 0.95\%                 \\ \cline{3-6} \cline{9-12}
                     &                          & Kohl's Corp.                                                     & Consumer Discretionary                                           & (0.135)             & (0.88\%)               &                      &                          & niversal Health Services, Inc.                                                  & Health Care            & 0.142               & 0.93\%                 \\ \cline{3-6} \cline{9-12}
                     &                          & Macy's Inc.                                                      & Consumer Discretionary                                           & (0.134)             & (0.87\%)               &                      &                          & SL Green Realty                                                                 & Financials             & 0.132               & 0.86\%                 \\ \cline{3-6} \cline{9-12}
                     &                          & Target Corp.                                                     & Consumer Discretionary                                           & (0.133)             & (0.86\%)               &                      &                          & Cardinal Health Inc.                                                            & Health Care            & 0.116               & 0.76\%                 \\ \cline{3-6} \cline{9-12}
                     &                          & L Brands Inc.                                                    & Consumer Discretionary                                           & (0.128)             & (0.83\%)               &                      &                          & Dentsply Sirona                                                                 & Health Care            & 0.113               & 0.74\%                 \\ \cline{3-6} \cline{9-12}
                     &                          & Baxter International Inc.                                        & Health Care                                                      & 0.126               & 0.82\%                 &                      &                          & Prologis                                                                        & Financials             & 0.110               & 0.72\%                 \\ \cline{3-6} \cline{9-12}
                     &                          & TJX Companies Inc.                                               & Consumer Discretionary                                           & (0.116)             & (0.75\%)               &                      &                          & Welltower Inc.                                                                  & Financials             & 0.110               & 0.72\%                 \\ \cline{3-6} \cline{9-12}
                     &                          & Urban Outfitters                                                 & Consumer Discretionary                                           & (0.111)             & (0.72\%)               &                      &                          & Patterson Companies                                                             & Health Care            & 0.109               & 0.71\%                 \\ \cline{3-6} \cline{9-12}
                     &                          & Vertex Pharmaceuticals Inc                                       & Health Care                                                      & 0.110               & 0.72\%                 &                      &                          & TECO Energy                                                                     & Utilities              & (0.109)             & (0.71\%)               \\ \cline{3-6} \cline{9-12}
                     &                          & Costco Co.                                                       & Consumer Staples                                                 & (0.107)             & (0.70\%)               &                      &                          & Humana Inc.                                                                     & Health Care            & 0.109               & 0.71\%                 \\ \hline
\multirow{10}{*}{8}  & \multirow{10}{*}{0.0116} & Aon plc                                                          & Financials Insurance                                             & (0.177)             & (1.17\%)               & \multirow{10}{*}{8}  & \multirow{10}{*}{0.0088} & Tegna                                                                           & Consumer Discretionary & 0.155               & 0.99\%                 \\ \cline{3-6} \cline{9-12}
                     &                          & Hartford Financial Svc.Gp.                                       & Financials Insurance                                             & (0.160)             & (1.06\%)               &                      &                          & Dollar Tree                                                                     & Consumer Discretionary & (0.148)             & (0.95\%)               \\ \cline{3-6} \cline{9-12}
                     &                          & Chubb Corp                                                       & Financials Insurance                                             & (0.153)             & (1.01\%)               &                      &                          & Wal-Mart Stores                                                                 & Consumer Discretionary & (0.143)             & (0.91\%)               \\ \cline{3-6} \cline{9-12}
                     &                          & Marsh \& McLennan                                                & Financials Insurance                                             & (0.147)             & (0.97\%)               &                      &                          & Ross Stores                                                                     & Consumer Discretionary & (0.129)             & (0.83\%)               \\ \cline{3-6} \cline{9-12}
                     &                          & Abbott Laboratories                                              & Health Care                                                      & 0.139               & 0.92\%                 &                      &                          & Celgene Corp.                                                                   & Health Care            & (0.121)             & (0.77\%)               \\ \cline{3-6} \cline{9-12}
                     &                          & ACE Limited                                                      & Financials Insurance                                             & (0.132)             & (0.87\%)               &                      &                          & Costco Co.                                                                      & Consumer Staples       & (0.114)             & (0.73\%)               \\ \cline{3-6} \cline{9-12}
                     &                          & American International Group, Inc.                               & Financials Insurance                                             & (0.130)             & (0.86\%)               &                      &                          & AutoZone Inc                                                                    & Consumer Discretionary & (0.113)             & (0.72\%)               \\ \cline{3-6} \cline{9-12}
                     &                          & Smucker (J.M.)                                                   & Consumer Staples                                                 & 0.130               & 0.86\%                 &                      &                          & Interpublic Group                                                               & Consumer Discretionary & 0.111               & 0.71\%                 \\ \cline{3-6} \cline{9-12}
                     &                          & Merck \& Co.                                                     & Health Care                                                      & 0.125               & 0.83\%                 &                      &                          & Gilead Sciences                                                                 & Health Care            & (0.111)             & (0.71\%)               \\ \cline{3-6} \cline{9-12}
                     &                          & CVS Health                                                       & Consumer Staples                                                 & 0.125               & 0.82\%                 &                      &                          & Kroger Co.                                                                      & Consumer Staples       & (0.110)             & (0.71\%)               \\ \hline
\multirow{10}{*}{9}  & \multirow{10}{*}{0.0104} & Unum Group                                                       & Financials Insurance                                             & (0.181)             & (1.26\%)               & \multirow{10}{*}{9}  & \multirow{10}{*}{0.0077} & Hormel Foods Corp.                                                              & Consumer Discretionary & 0.166               & 1.08\%                 \\ \cline{3-6} \cline{9-12}
                     &                          & United Health Group Inc.                                         & Health Care                                                      & (0.170)             & (1.18\%)               &                      &                          & Vertex Pharmaceuticals Inc                                                      & Health Care            & (0.145)             & (0.94\%)               \\ \cline{3-6} \cline{9-12}
                     &                          & Marsh \& McLennan                                                & Financials Insurance                                             & (0.167)             & (1.16\%)               &                      &                          & PPL Corp.                                                                       & Utilities              & (0.128)             & (0.83\%)               \\ \cline{3-6} \cline{9-12}
                     &                          & ACE Limited                                                      & Financials Insurance                                             & (0.164)             & (1.14\%)               &                      &                          & Kellogg Co.                                                                     & Consumer Staples       & 0.123               & 0.80\%                 \\ \cline{3-6} \cline{9-12}
                     &                          & XL Capital                                                       & \begin{tabular}[c]{@{}c@{}}Financials Insurance\end{tabular} & (0.158)             & (1.09\%)               &                      &                          & Alexion Pharmaceuticals                                                         & Pharmaceutical         & (0.121)             & (0.79\%)               \\ \cline{3-6} \cline{9-12}
                     &                          & Hartford Financial Svc.Gp.                                       & Financials Insurance                                             & (0.158)             & (1.09\%)               &                      &                          & Bed Bath \& Beyond                                                              & Consumer Discretionary & (0.119)             & (0.77\%)               \\ \cline{3-6} \cline{9-12}
                     &                          & Aon plc                                                          & Financials Insurance                                             & (0.156)             & (1.08\%)               &                      &                          & Pulte Homes Inc.                                                                & Consumer Discretionary & (0.118)             & (0.77\%)               \\ \cline{3-6} \cline{9-12}
                     &                          & CIGNA Corp.                                                      & Health Care                                                      & (0.149)             & (1.03\%)               &                      &                          & American Electric Power                                                         & Utilities              & (0.117)             & (0.76\%)               \\ \cline{3-6} \cline{9-12}
                     &                          & Humana Inc.                                                      & Health Care                                                      & (0.132)             & (0.91\%)               &                      &                          & Celgene Corp.                                                                   & Health Care            & (0.115)             & (0.75\%)               \\ \cline{3-6} \cline{9-12}
                     &                          & BB\&T Corporation                                                & \begin{tabular}[c]{@{}c@{}}Financials Insurance\end{tabular} & 0.131               & 0.91\%                 &                      &                          & Gap (The)                                                                       & Consumer Discretionary & 0.114               & 0.74\%                 \\ \hline
\multirow{10}{*}{10} & \multirow{10}{*}{0.009}  & Sysco Corp.                                                      & Consumer Staples                                                 & 0.177               & 1.17\%                 & \multirow{10}{*}{10} & \multirow{10}{*}{0.0074} & Alexion Pharmaceuticals                                                         & Pharmaceutical         & 0.164               & 1.04\%                 \\ \cline{3-6} \cline{9-12}
                     &                          & St Jude Medical                                                  & Health Care                                                      & (0.152)             & (1.00\%)               &                      &                          & Gap (The)                                                                       & Consumer Discretionary & (0.145)             & (0.92\%)               \\ \cline{3-6} \cline{9-12}
                     &                          & The Hershey Company                                              & Consumer Staples                                                 & 0.152               & 1.00\%                 &                      &                          & Interpublic Group                                                               & Consumer Discretionary & (0.135)             & (0.86\%)               \\ \cline{3-6} \cline{9-12}
                     &                          & PepsiCo Inc.                                                     & Consumer Staples                                                 & 0.151               & 1.00\%                 &                      &                          & Union Pacific                                                                   & Industrials            & 0.132               & 0.84\%                 \\ \cline{3-6} \cline{9-12}
                     &                          & Northrop Grumman Corp.                                           & Industrials                                                      & 0.142               & 0.94\%                 &                      &                          & Ross Stores                                                                     & Consumer Discretionary & (0.131)             & (0.83\%)               \\ \cline{3-6} \cline{9-12}
                     &                          & Celgene Corp.                                                    & Health Care                                                      & (0.138)             & (0.91\%)               &                      &                          & Harman Int'l Industries                                                         & Consumer Discretionary & (0.128)             & (0.81\%)               \\ \cline{3-6} \cline{9-12}
                     &                          & Expeditors Int'l                                                 & Industrials                                                      & (0.132)             & (0.87\%)               &                      &                          & AutoZone Inc                                                                    & Consumer Discretionary & (0.116)             & (0.74\%)               \\ \cline{3-6} \cline{9-12}
                     &                          & C. H. Robinson Worldwide                                         & Industrials                                                      & (0.128)             & (0.85\%)               &                      &                          & Autodesk Inc                                                                    & Information Technology & 0.116               & 0.74\%                 \\ \cline{3-6} \cline{9-12}
                     &                          & Reynolds American Inc.                                           & Consumer Staples                                                 & 0.127               & 0.84\%                 &                      &                          & Precision Castparts Corp.                                                       & Aerospace and defense  & 0.116               & 0.74\%                 \\ \cline{3-6} \cline{9-12}
                     &                          & Johnson Controls                                                 & Consumer Discretionary                                           & 0.126               & 0.84\%                 &                      &                          & Smucker (J.M.)                                                                  & Consumer Staples       & (0.115)             & (0.73\%)               \\ \hline
\end{tabular}}
\caption[Eigenvector components during 2004-2005 and 2008-2009]{Eigenvectors corresponding to top 10 largest eigenvalues are displayed. For each eigenvector, largest (in absolute value) components are listed.  The contribution of each firm in the first eigenvector is uniform, which implies the market mode. Other eigenvectors represent business sectors.}
\label{table::eigenvector}
\end{table}

\subsection{Dynamic experiment: implications of $\hat b$}
\begin{figure}[!htbp]
\centering
\begin{tabular}{c}
\includegraphics[width=0.8\linewidth]{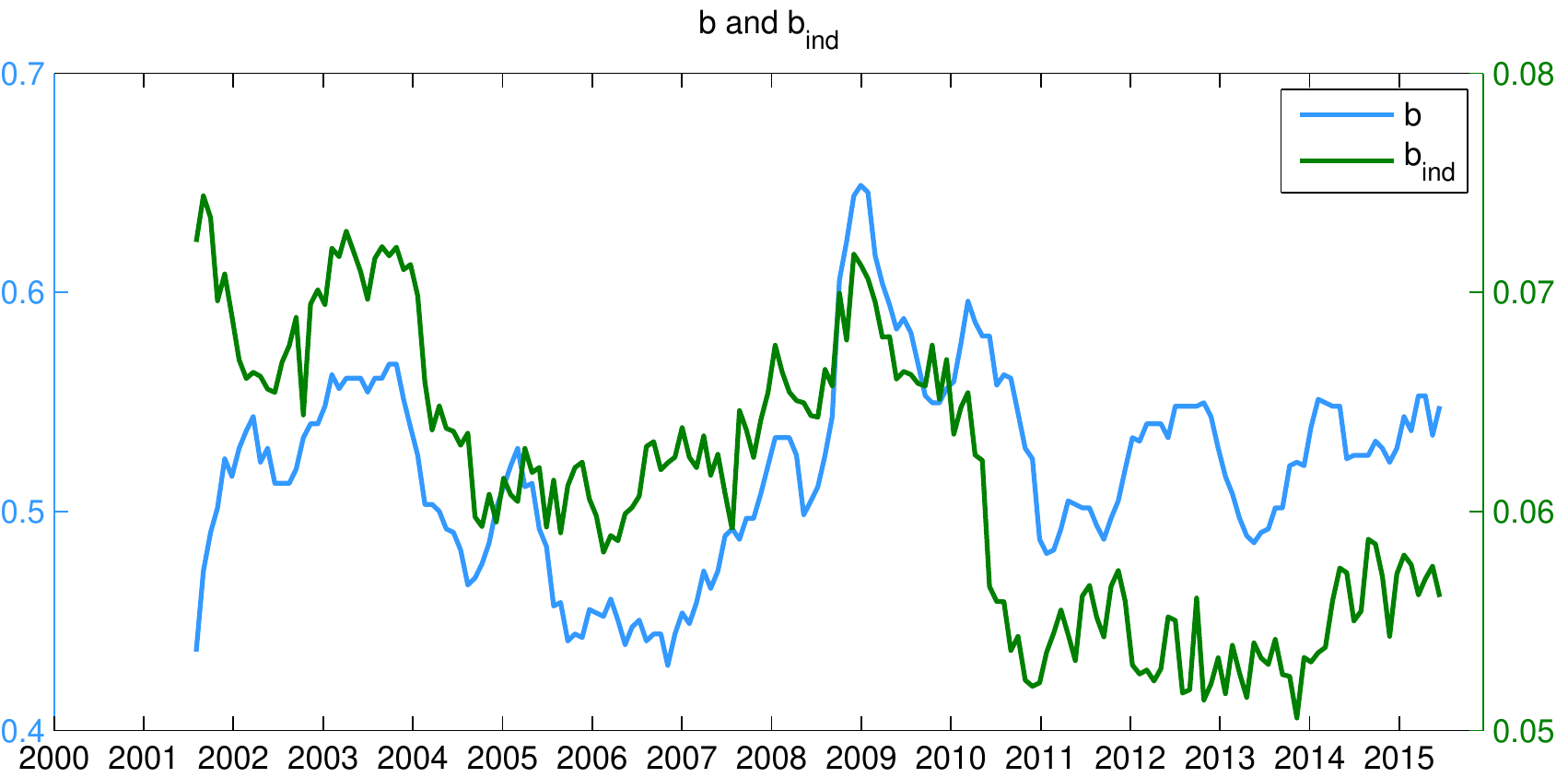}
\end{tabular}
\caption[Our estimator for $b$ and the average of estimated individual AR(1) coefficients]{Our estimator for $b$ and the average of estimated individual AR(1) coefficients (denoted as $b_{ind}$ in the plot). Although there is a scale difference between the two quantities, they behave very similarly, showing slower mean-reversions of residual returns in crisis.}
\label{fig::timeseries_MRT}
\end{figure}

We now concentrate on $\hat b$, the estimator for the ``mean-field'' or ``representative'' autoregressive coefficient of residuals. To illustrate the meaning of $\hat b$, we compare it with the behavior of $\hat b_i$ for each residual, where $b_i$ is the estimated AR(1) coefficient for $i$-th residual, such that $U_{it} = b_i U_{i,t-1} + \epsilon_{it}$. Let us define an estimator $\widehat{b_{ind}} \defeq  \frac{1}{N}\sum\limits_{i=1}^N \hat {|b_i|}$ \footnote{Here we take the absolute value for $b_i$, since the limiting spectrum for vector AR(1) processes depends only on the magnitude of coefficients (see Eq. \ref{eq::poly})}. Figure \ref{fig::timeseries_MRT} plots the evolutions of $\hat b$ and $\widehat{b_{ind}}$. Except the scale difference\footnote{Note that the scale of the discrepancy may be due to the fact that $\hat b$ coarsely integrates many outliers or complicated correlation cases.}, the overall patterns of the two quantities are very similar. They both indicate that during crisis, the residual returns tend to be trending (i.e., having longer mean-reversion times) than normal periods. To check whether these patterns are generic for residuals, we also estimated AR(1) coefficients from each \emph{original} return. We found that although it also increases in crisis, its behavior is not close to $\hat b$ compared to that of residuals. Therefore, in the context of Section \ref{sec::mft}, $\hat b$ is the bulk coefficient that delivers compressed information from of the coefficients of all residuals.

\subsection{Market dynamics from estimators}
\begin{figure}[!htbp]
\centering
\begin{tabular}{c}
\includegraphics[width=0.7\linewidth]{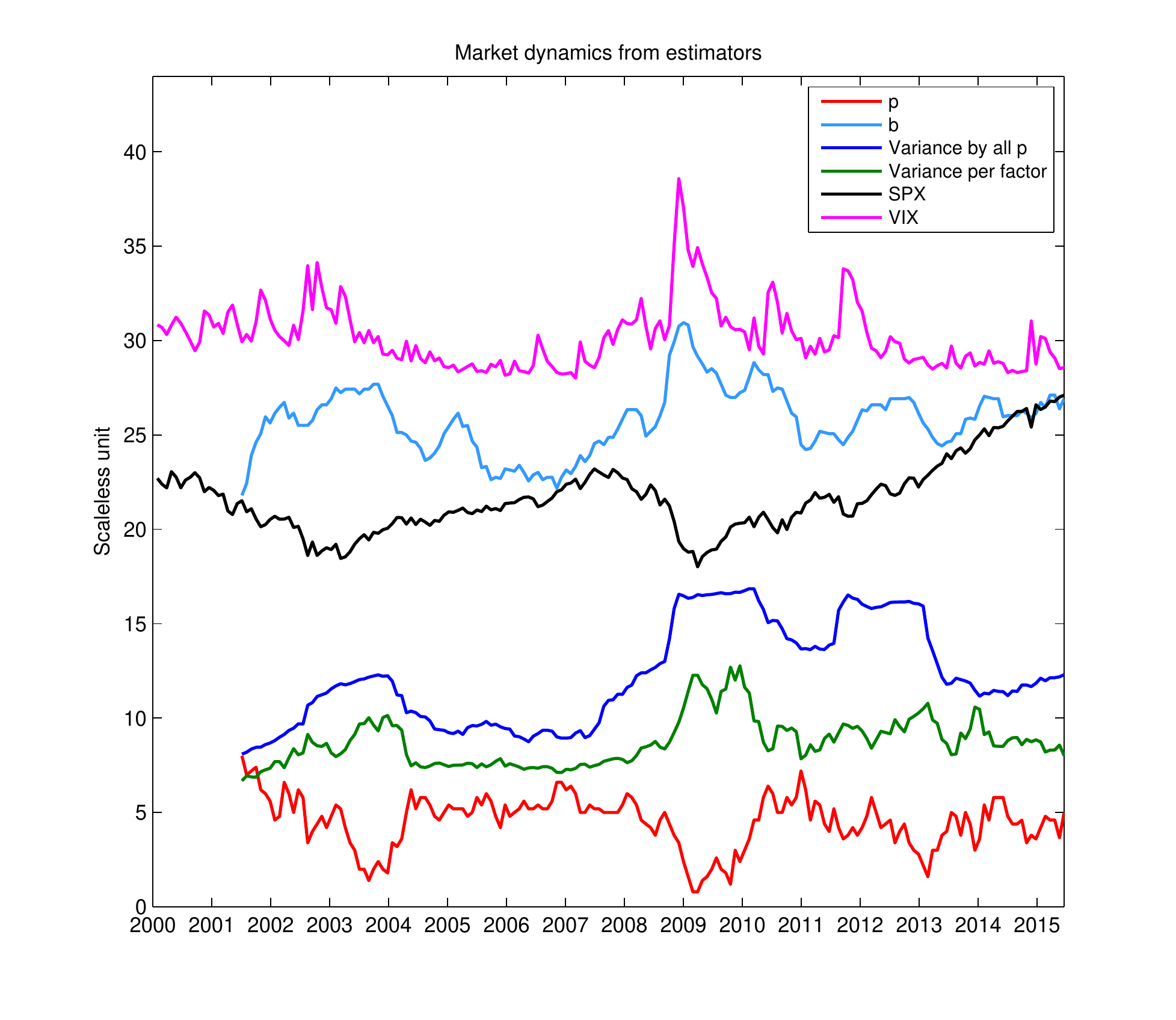}
\end{tabular}
\caption[Market dynamics captured by estimators and variance explanations]{Market dynamics captured by estimators and variance explanations: $\hat p$, $\hat b$, variance explained by $\hat p$, variance explained per factor, S\&P500 index (SPX), and volatility index (VIX). Each quantity is obtained with moving windows, and rescaled for comparisons. It is clear that the variance of the whole market is the most condensed during the crisis (2008-2009). In addition, the overall trends after the crisis is different, showing sporadic market condensations. The number of investigated stocks is $N = 378$ and the time window is also set to be the same length, $T$  = 378 business days ($\sim$ 1.5 years). Note that the estimated values are available only from mid-2001, because of the length of moving window for estimations.}
\label{fig::timeseries_1}
\end{figure}

As seen from previous discussions, the estimated parameters from factor models provide informative guidance on market dynamics. Figure \ref{fig::timeseries_1} displays the evolution of our estimators, along with other market indicators such as equity market index (SPX) and volatility index (VIX). Note that all quantities are closely related to each other. For example, the estimator for autoregressive coefficient of residuals, $\hat b$, reflects the market movement. Most of time, $\hat b$ is mimicking the behaviors of VIX. Thus, this estimator reflect essential information on market fluctuations. In addition, We calculate the variance explained by $\hat p$ factors and the variance per factor. The estimated number of factors sharply decreases in the crisis (2008-2009). At the same period, the variance explained per factor is sharply increasing, indicating the market condensation phenomenon. That is, during the major market events in 2008, correlations changed dramatically, even affecting previously uncorrelated sectors. Thus, the whole market moves together, which increases the largest eigenvalues of the correlation matrix.

\section{Conclusions}
\label{sec::conclusions}
Random matrix theory is gaining increasing attentions for analyzing complex high-dimensional data. This paper relates the factor model estimation problem to fitting empirical eigenvalue distribution of the covariance matrix. The spectrum from real data is complex and cannot be trivially dissected by traditional usage of the Marchenko-Pastur law or mere counting of the largest eigenvalues. Instead, we present a new approach to estimate factor models, by allowing control for both the number of factors and the correlation structure of residuals. Under reasonable assumptions for approximate factor models,  we show how our estimation problem is applied in high-dimension settings. In addition, by using the free random variable techniques and modified estimation problem, the implementation of our method is done efficiently. Monte Carlo analysis shows that the proposed method boosts up the power of identification of weak factors and that the performance is less affected by signal-to-noise ratios. Furthermore, from the application to real data with moving windows, we monitor how our estimators effectively characterize the market dynamics.

Several future studies are planned. Clearly, further research will be needed to employ the more general residual modeling, for which we can calculate the distribution readily. For example, as described in \cite{Burda2010}, if we consider vector ARMA(1,1) processes, we have up to 6th-order polynomial equations. A possible extension is to develop a more delicate method to dynamically estimate the residual covariance matrix, so that the residual processes can be exploited for more practical purposes, such as mean-reversion dependence structures in large dimensions. There is an interesting connection between our study to the covariance matrix estimations. The covariance matrix estimations via factor models have been investigated by \cite{Fan2008, Fan2011, Fan2013}. To apply our method to different frequency data would be also interesting.

\section{Acknowledgement}
JY gratefully acknowledges the support of ILJU Foundation Scholarship.

\bibliography{bibliography_yeo}

\appendix
%\begin{appendix}

\section{Preliminaries for spectrum-based estimations}
\label{sec::preliminaries}
This section provides preliminaries that are required for the supporting theory to our estimation method.

\begin{definition}[Empirical spectral distribution]
\label{def::1}
Let $A_n$ be an $n\times n$ matrix having real eigenvalues  $\lambda_1\leq \cdots \leq \lambda_n$. Then the empirical spectral distribution of $A_n$ is defined as
\begin{eqnarray}
\mathcal{F}^{A_n} (x) &=& \cfrac{1}{n}\sum\limits_{i=1}^n \mathbf{1}_{\{\lambda_i(A_n) \leq x \}}
\end{eqnarray}
where $\mathbf{1}_{\{\cdot\}}$ denotes the indicator function of the set $\{\cdot\}$.
\end{definition}
\begin{definition}[The Stieltjes transform]
\label{def::2}
Let $\mathcal{F}(x)$ be any function of bounded variation. Then the Stieltjes transform of $\mathcal{F}(x)$ is defined as
\begin{eqnarray}
m_F(z) = \int\cfrac{1}{x-z}d\mathcal{F}(x), \ \ (Im (z) > 0)
\end{eqnarray}
\end{definition}

\begin{assumption}\label{as::1}
The general covariance matrix $C_N$ has the form
\begin{eqnarray}
C_N = \cfrac{1}{T}A_N^{1/2}\epsilon B_T\epsilon^TA_N^{1/2}
\end{eqnarray}
where $\epsilon$ is an $N \times T$ random matrix with i.i.d. entries, and $A_N$ and $B_T$ are deterministic symmetric semi-definite matrices of size $N\times N$ and $T \times T$, respectively.
\end{assumption}
\begin{assumption}\label{as::2}
T = T(N) and there exists a positive constant $c$ such that
\begin{eqnarray}
\frac{N}{T(N)} \rightarrow c, \ \ as \ \ N\rightarrow \infty,
\end{eqnarray}
\end{assumption}
\begin{assumption}\label{as::3}
$\epsilon_{it}$ are i.i.d and $\mathbf{E}\epsilon_{it} = 0$, $\mathbf{E}|\epsilon_{it}|^2 = 1$,  $\mathbf{E}|\epsilon_{it}|^4 < \infty$.
\end{assumption}
\begin{assumption}\label{as::4}
$\mathcal{F}^{A_N}$ and $\mathcal{F}^{B_T}$ weakly converge to non-random probability density functions $\mathcal{F}^{A}$ and $\mathcal{F}^B$, as $N \rightarrow \infty$.
\end{assumption}
\begin{assumption}\label{as::5}
$\parallel A_N \parallel$ and $\parallel B_T \parallel$ , the respective spectral norms of $A_N$ and $B_T$, are bounded in N.
\end{assumption}

Note that the class of matrices of the form $C_N$ in Assumption \ref{as::1} appears in various applications, such as multiple-input multiple-output (MIMO) system in wireless communications or in financial time series where $A_N$ and $B_T$ represent the cross- and serial- correlation structure of data. This class of model is also known as the separable covariance model, since there is no space-time interaction. As discussed in this paper, the motivation of this assumption is natural, as the approximate factor model allows cross-sectional and serial correlations in residuals. The Assumption \ref{as::2} requires that $N$ and $T$ are comparable asymptotically. Assumption \ref{as::3} indicates moment conditions, so that the maximum eigenvalues of $\frac{1}{T}\epsilon\epsilon^T$ do not diverge. Assumption \ref{as::4} restates the convergence of the empirical spectral distribution to non-random limiting distributions. Assumption \ref{as::5} restricts unusual large variations of idiosyncratic components.

Now here we state the main result of \cite{Zhang2006}.
\begin{lemma} [\cite{Zhang2006}]
\label{lem::1}
If Assumptions \ref{as::1} to \ref{as::4} hold, the eigenvalue distribution of $C_N = \cfrac{1}{T}A_N^{1/2}\epsilon B_T\epsilon^TA_N^{1/2}$ converges weakly to a non-random distribution $\mathcal{F}^{{c,A,B}}$. The Stieltjes transform of $\mathcal{F}^{{c,A,B}}$, $m(z)$, together with other analytical function $p(z)$ and $q(z)$, constitutes a solution to the system

\systeme{
m(z) = -z^{-1}(1-c) - z^{-1}c\int\cfrac{1}{1+q(z)x}d\mathcal{F}^A(x),
m(z) = -z^{-1}\int\cfrac{1}{1+p(z)y}d\mathcal{F}^B(y),
m(z) = -z^{-1}-p(z)q(z)}\\ \\
which is unique in the set $\{(m(z), p(z), q(z)): Im (m(z)) > 0, Im (p(z)) > 0, Im (q(z))$ $> 0 \}$.
\end{lemma}
%This lemma elegantly presents the convergence of the empirical spectral distributions of the general sample covariance matrices and identified the limiting spectral distribution through a system of equations determining its Stieltjes transform. Our method is based on this. Once we assume residuals have doubly-correlated structure, such that $U=A_N^{1/2}\epsilon B_T^{1/2}$, this lemma asserts the convergence of empirical spectrum of residuals. However, we will not solve this system of equations, since the analytic expression with general form of $A_N$ and $B_T$ is implicit as well as very complicated to deal with. Rather, we introduce an approximate model with simple parameterizations, and directly derive the probability distribution of eigenvalues by using the techniques introduced in \cite{Burda2010}. Then we relate the spectrum of the model to real data.

The boundedness of eigenvalues in the support of $\mathcal{F}^{c,A_N,B_T}$ is known as shown in the following lemma.
\begin{lemma} [\cite{PS2009}]
\label{lem::2}
Suppose Assumption \ref{as::1} to \ref{as::5} hold. Let $\epsilon$ have Gaussian entries, or either $A_N$ or $B_T$ be a diagonal matrix . Then \\

$\mathbf{P}$(no eigenvalue of $C_N$ appears in $[a,b]$ for all large $N$) = 1\\ \\
where the interval $[a,b]$ with $a>0$ lies in an open interval outside the support of $\mathcal{F}^{c,A_N,B_T}$.
\end{lemma}
\begin{definition}[Factor models]
A factor model  for $N$ assets and $T$ observations is written as
\begin{eqnarray}
\label{eq::factormodel}
R = LF + U
\end{eqnarray}
where $R$ is an $N\times T$ matrix of data, $p$ is the number of factors, $L$ is an $N \times p$ matrix of factor loadings, $F$ is a $p \times T$ matrix of factors, and $U$ is an $N \times T$ matrix of the idiosyncratic components of residuals.
\end{definition}
The rationale of this factor model is to linearly decompose the original signal into systemic components (factors) and idiosyncratic components (residuals).
\begin{definition}[Principal components]
For any $p$, the methods of principal components minimizes
\begin{eqnarray}
&&\min\limits_{L,F}(NT)^{-1} \|R - LF \|_{\texttt{Frob}}\\
&&s.t. \ \ \frac{1}{T}FF^T = I_p \ \ or \ \ \frac{1}{N}L^L = I_p
\end{eqnarray}
\end{definition}
One solution for the above problem is given as
\begin{eqnarray}
\widehat{L} &=& \sqrt {N} \times (\texttt{eigenvectors corresponding to} \\
&& \texttt{the } p \texttt{ largest eigenvalues of } R^TR) \\
\widehat{F} &=& \frac {1}{N}\widehat{L}^TR
\end{eqnarray}
Note that as $N,T\rightarrow \infty$, common components $LF$ can be consistently estimated by $\widehat{L}\widehat{F}$  \cite{Bai2003, Fan2013}.
\begin{consistency}[Factor model estimations]
\label{consistency::1}
Suppose the assumptions \ref{as::1}--\ref{as::5} hold and further assume that covariance of $U$ are separable and the cross- and auto-covariance matrix of $U (= R - LF)$ are given as $A_N^*$ and $B_T^*$, which are parameterized by $\theta_{A_N^*}$ and $\theta_{B_T^*}$, respectively. Suppose that the true number of factors is $p^*$, and that common components $LF$ are consistently estimated by $\hat L \hat F$ which is obtained from the method of principal components, then the estimators in Eq. \ref{eq::problem}, $\{\hat p, \hat \theta_{A_N}, \hat \theta_{B_T}\}$, converge to  $\{p^*, \theta_{A_N^*}, \theta_{B_T^*}\}$, as $N,T\rightarrow \infty$.
\end{consistency}

\section{A brief overview of free random variables techniques}
\label{sec::frv}
\subsection{Key concepts}
In this section, we summarize main concepts and key results of the technique that we employed. We will follow the notations and derivations from \cite{Burda2005} and \cite{Burda2010}. Throughout this section, we assume a simple decomposition of covariance structures
\begin{eqnarray*}
Cov_{ia,jb} = A_{ij}B_{ab}
\end{eqnarray*}
$i,j=1,\dots, N$, $a,b=1,\dots, T$, $A$ is a $N\times N$ cross-covariance matrix and $B$ is a $T\times T$ auto-covariance matrix.
Suppose $\epsilon$ is $N\times T$ uncorrelated Gaussian random matrix. Then a correlated Gaussian random matrix $U$ (e.g., $N\times T$ time series) can be written as
\begin{eqnarray*}
U = A^{1/2} \epsilon B^{1/2}
\end{eqnarray*}
Define the sample (empirical) covariance matrix $C$ as
\begin{eqnarray*}
C = \frac{1}{T}UU^T, \ \ \ \
\end{eqnarray*}
We will show the relation between $C$ and $A$, $B$, using free random variable techniques. It generalizes the results for the eigenvalue density of large-dimensional empirical covariance matrices with doubly-correlated structure.

First, note that the relationship between empirical spectral density ($\rho_{H}(\lambda)$) and Green's function ($G_{H}(z)$) is the following:
\begin{eqnarray*}
\rho_{H}(\lambda) = -\frac{1}{\pi}\lim\limits_{\epsilon \rightarrow 0^+} \mathfrak{Im} G_{H} (\lambda + i\epsilon).
\end{eqnarray*}
This Green's function generates moments of a probability distribution, where the $n$-th moment is defined by
\begin{definition}[Moment]
\begin{eqnarray*}
m_n &=& \frac{1}{N} \langle \texttt{Tr} H^n \rangle \\
&=& \int \rho_H (\lambda) \lambda^n d\lambda
\end{eqnarray*}
\end{definition}
\begin{definition}[Moment generating function]
\begin{eqnarray*}
G_{H}(z) &=& \sum\limits_{n\geq 0} \frac{m_n}{z^{n+1}}\\
M_{H}(z) &=& \sum\limits_{n\geq 1} \frac{m_n}{z^{n+1}}
\end{eqnarray*}
\end{definition}
This suggest the the relation between $G_{H}(z)$ and $M_{{H},n}$ as
\begin{eqnarray*}
M_{{H}}(z) = z G_{{H}}(z) - 1.
\end{eqnarray*}
There is the inverse transform of the Green's function and moment generating function.
\begin{definition}[Blue's function and N-transform]
\begin{eqnarray*}
G_{{H}}(B_{H}(z)) = B_{H}(G_{H}(z)) = z \\
M_{{H}}(N_{H}(z)) = N_{H}(M_{H}(z)) = z
\end{eqnarray*}
\end{definition}

Now we return to our original objective, empirical covariance matrix, $C$. Recall that it can be expressed as
\begin{eqnarray*}
C &=& \frac{1}{T}UU^T\\
 &=& \frac{1}{T} {{A}^{1/2} \epsilon B \epsilon^T {A}^{1/2}}
\end{eqnarray*}
For arbitrary $A$ and $B$, the $N$-transform of $C$ can be derived as
\begin{eqnarray*}
N_{C}(z) &=& N_{\frac{1}{T} {{A}^{1/2} \epsilon B \epsilon^T {A}^{1/2}}}(z) \\
&=& N_{\frac{1}{T} {\epsilon B \epsilon^T {B}}}(z) \ \ \because \emph{cyclic property of trace}\\
&=& \frac{z}{1+z} N_{\frac{1}{T} \epsilon B \epsilon^T }(z) N_{A} (z) \ \ \because \emph{FRV multiplication law} \\
&=& \frac{z}{1+z} N_{\frac{1}{T} \epsilon^T\epsilon B  }(rz)N_B (z)  \ \ \because \emph{cyclic property of trace + rescaling} \\
&=& \frac{z}{1+z} \frac{rz}{1+rz} N_{\frac{1}{T} \epsilon^T \epsilon }(rz) N_{B}(rz) N_{A} (z)  \ \ \because \emph{FRV multiplication law} \\
&=& rz N_{B}(rz)N_{A} (z) \ \ \because N_{\frac{1}{T}\epsilon^T\epsilon}(z) = \frac{(1+z)(r+z)}{z}
\end{eqnarray*}
Using the moments' generating function $M \equiv M_C(z)$ and its inverse relation to $N$-transform, this can be written as,
\begin{eqnarray*}
N_{C}(z) &=& rz N_{B}(rz)N_{A} (z) \\
\Longleftrightarrow \ \ \ \ z &=& rMN_{B}(rM)N_{A}(M)
\end{eqnarray*}
We will use this equation to compute the spectral density for given matrix $A$ and $B$.
\subsection{The case of our simplied model: $U_{it} = b U_{i,t-1} + \xi_{it}$}
Suppose $U_{it}$ ($n = 1, \cdots, N$, $t = 1, \cdots, T$) be a time-series,  following the autoregressive model:
\begin{eqnarray*}
U_{it} = b U_{i,t-1} + \xi_{it}
\end{eqnarray*}
where $|b|<1$ and $\xi_{it} \sim N(0, 1-b^2)$. Let $c = \frac N T$.
The free random variables technique provide analytic derivation for the eigenvalue distribution of correlation matrix $C = \frac 1 T U U^T$.

Our goal is to find $\rho_C(\lambda)$. The strategy is the following:
\begin{enumerate}
\item
Find $M_C(z)$, from the equation for $N$-transform.
\item
Find $G_C(z)$, by $M_{C}(z) = z G_{C}(z) - 1$.
\item
Find $\rho_C(\lambda)$, by $\rho_C(\lambda) = -\frac{1}{\pi}\lim\limits_{\epsilon \rightarrow 0^+} \mathfrak{Im} G_{C} (\lambda + i\epsilon)$.
\end{enumerate}
Other than the first part is straightforward, so let us examine the equation for $N$-transform.

Recall that we have the equation for $N$-transform for arbitrary matrix $A$ and $B$ as
\begin{eqnarray*}
z = rMN_{B}(rM)N_{A}(M)
\end{eqnarray*}
For the vector AR(1) process considered above, the cross-correlation matrix $A = I_{N}$. Then $N_{A}(z) = N_{I}(z) = 1 + 1/z$. Thus, the above equation can be rewritten as
\begin{eqnarray*}
z &=& rMN_{B}(rM)(1 + 1 / M) = r (1+M) N_{B}(rM) \\
\Longleftrightarrow \frac{z}{r(1+M)} &=& N_B(rM) \\
\label{eq::ntransform}
\Longleftrightarrow rM &=& M_B\Bigg(\frac{z}{r(1+M)}\Bigg).
\end{eqnarray*}

Now we will need to find $M_B$. The two-point covariance function for VAR(1) is the following.
Note that the auto-covariance matrix of vector AR(1) process we consider has a simple form:
\begin{eqnarray*}
B_{st} = \frac{\var(\xi)}{1-b^2}b^{|s-t|} = b^{|s-t|} \ \ \because \var(\xi)=1-b^2.
\end{eqnarray*}
Using Fourier-transform of the matrix $B$, it can be shown that the moment generating function of $B$ is give by
\begin{eqnarray*}
M_B(z) = -\frac{1}{\sqrt{1-z}\sqrt{1-\frac{(1+b^2)^2}{1-b^2}z}}.
\end{eqnarray*}
Now we solve Eq. \ref{eq::ntransform} for $M_B$, which leads to the following polynomial equation (with $a^2=1-b^2$):
\begin{eqnarray*}
&&a^4c^2M^4 + 2a^2c(-(1+b^2)z+a^2c)M^3 + ((1-b^2)^2z^2-2a^2c(1+b^2)z \\
&&+ (c^2-1)a^4)M^2-2a^4M-a^4=0
\end{eqnarray*}
Thus, we obtain the first step. The other steps are followed straightforwardly.
\section{Numerical calculation of Kullback-Leibler divergence}
\label{app::sd}
The spectral distance measure we use requires the calculation of Kullback-Leibler (KL) divergence.
\begin{eqnarray*}
\mathcal{D}_{KL}(P\|Q) = \sum\limits_i P_i \log \frac{P_i}{Q_i}
\end{eqnarray*}
where $P$ and $Q$ are probability densities, and $P_i = P(ih)$ with grid size $h$. To deal with zero elements of $P_i$ that possibly appear due to the spectral characteristics of empirical covariance matrix, we use $\widetilde{P}_i$ from the following manipulation. For a small $\varepsilon>0$,
\begin{eqnarray*}
\widetilde{P}_i &=&
\begin{cases}
      \alpha P_i, & \ \ {\emph{if }P_i > 0 }\\
      \varepsilon, & \ \ {\emph{if }P_i = 0 }
    \end{cases} \\
where && \\
\alpha &=& 1 - (\emph{number of zeros of }P_i)\varepsilon
\end{eqnarray*}
where we use the fact that $\sum_i \widetilde{P}_i = 1$ and $\varepsilon$ is assumed to be small enough such that $\varepsilon \ll 1/(\emph{number of zeros of }P_i)$.

\end{document}